\documentclass{elsart}

\usepackage{amssymb}
\usepackage{graphicx}

\begin{document}

\begin{frontmatter}



\title{Triaxiality in the interacting boson model}


\author[Caen,Istanbul]{B.~Sorgunlu}
and
\author[Caen]{P.~Van~Isacker}

\address[Caen]{Grand Acc\'el\'erateur National d'Ions Lourds,
CEA/DSM--CNRS/IN2P3, BP~55027, F-14076 Caen Cedex 5, France}
\address[Istanbul]{Department of Physics, Istanbul University,
Istanbul, Turkey}

\begin{abstract}
The signature splitting of the $\gamma$-vibrational band
of several Ru, Pd, Xe, Ba, Os and Pt isotopes
is analyzed in the framework of the interacting boson model (IBM).
The nuclei studied are close to the $\gamma$-unstable SO(6) limit of the IBM
and have well-known $\gamma$ bands.
It is shown that in most nuclei
the signature splitting is better reproduced
by the inclusion of a three-body interaction between the $d$ bosons.
In none of the nuclei evidence
for a stable, triaxial ground-state shape is found. 
\end{abstract}
\end{frontmatter}

\section{Introduction}
\label{s_intro}
Quadrupole deformations of atomic nuclei can be of two types:
$\beta$ deformations which preserve axial symmetry
and $\gamma$ deformations which lead to triaxial shapes.
Although the collective model of nuclei introduced these ideas
more than 50 years ago~\cite{Bohr53},
it is still a matter of debate to what extent triaxiality is present in nuclei
and, specifically, whether nuclear ground states exhibit stable triaxial deformation.
On the other hand, $\gamma$ bands,
associated with collective vibrations that break axial symmetry,
are a well-established feature in the spectroscopy of deformed nuclei.

In this paper we undertake a detailed analysis of $\gamma$-band properties,
in particular their signature splitting or even--odd staggering,
in order to shed light on the question of triaxiality.
The gist of the argument for doing this is as follows.
Theoretical spectra derived from potentials
which are either soft or rigid in the $\gamma$ degree of freedom,
display $\gamma$ bands with different signature splitting.
The observed energies of the $\gamma$-band members
can therefore be used to fix the degree of softness or rigidity in $\gamma$.
This idea is worked out here in the context
of the interacting boson model (IBM)~\cite{Iachello87}
which proposes an algebraic or group-theoretical description
of collective quadrupole excitations in nuclei.
The IBM is particularly well suited for our present purpose for two reasons.
First, its phenomenological application generally
leads to accurate calculations of nuclear properties at low energy.
As will be shown, a very precise description
of the $\gamma$-band signature splitting can be obtained
with three-body interactions between the bosons.
Second, once the algebraic hamiltonian is fitted to the data,
an intuitive, geometric picture in terms of $\beta$ and $\gamma$ deformations
can be obtained in the so-called classical limit.
In this way we establish a unbiased procedure
which gauges the importance of triaxiality from the data.

In the present study we focus our attention on nuclei
that are usually interpreted as soft in $\gamma$
[or close to the SO(6) limit of the IBM]
and we investigate to what extent
the observed signature splitting in the $\gamma$ band
signals the occurrence of more rigid triaxiality.

\section{The interacting boson model}
\label{s_mod}
In this section we give a brief description of the IBM
with particular emphasis on the version of the model
which includes higher-order interactions between the bosons.
A full account of the IBM is given in ref.~\cite{Iachello87}.

\subsection{The general hamiltonian}
\label{ss_general}
The building blocks of the IBM
are $s$ and $d$ bosons with angular momenta $\ell=0$ and $\ell=2$.
A nucleus is characterized by a constant total number of bosons $N$
which equals half the number of valence nucleons
(particles or holes, whichever is smaller).
In this paper no distinction is made between neutron and proton bosons,
an approximation which is known as \mbox{IBM-1}.

Since the hamiltonian of the \mbox{IBM-1} conserves the total number of bosons,
it can be written in terms of the 36 operators $b_{\ell m}^\dag b_{\ell' m'}$
where $b_{\ell m}^\dag$ ($b_{\ell m}$) creates (annihilates)
a boson with angular momentum $\ell$ and $z$ projection $m$.
It can be shown~\cite{Iachello87}
that this set of 36 operators generates the Lie algebra U(6)
of unitary transformations in six dimensions.
A hamiltonian that conserves the total number of bosons
is of the generic form
\begin{equation}
\hat H=E_0+\hat H^{(1)}+\hat H^{(2)}+\hat H^{(3)}+\cdots,
\label{e_ham}
\end{equation}
where the index refers to the order of the interaction
in the generators of U(6).
The first term $E_0$ is a constant
which represents the binding energy of the core. 
The second term is the one-body part
\begin{equation}
\hat H^{(1)}=
\epsilon_s[s^\dag\times\tilde s]^{(0)}+
\epsilon_d\sqrt{5}[d^\dag\times\tilde d]^{(0)}\equiv
\epsilon_s\hat n_s+
\epsilon_d\hat n_d,
\label{e_ham1}
\end{equation}
where $\times$ refers to coupling in angular momentum
(shown as an upperscript in round brackets),
$\tilde b_{\ell m}\equiv(-)^{\ell-m}b_{\ell,-m}$
and the coefficients $\epsilon_s$ and $\epsilon_d$
are the energies of the $s$ and $d$ bosons.
The third term in the hamiltonian~(\ref{e_ham})
represents the two-body interaction
\begin{equation}
\hat H^{(2)}=
\sum_{\ell_1\leq\ell_2,\ell'_1\leq\ell'_2,L}
\tilde v^L_{\ell_1\ell_2\ell'_1\ell'_2}
[[b^\dag_{\ell_1}\times b^\dag_{\ell_2}]^{(L)}\times
[\tilde b_{\ell'_2}\times\tilde b_{\ell'_1}]^{(L)}]^{(0)}_0,
\label{e_ham2}
\end{equation}
where the coefficients $\tilde v$ are related to the interaction matrix elements
between normalized two-boson states, 
\begin{equation}
\langle\ell_1\ell_2;LM|\hat H^{(2)}|\ell'_1\ell'_2;LM\rangle=
\sqrt{\frac{(1+\delta_{\ell_1\ell_2})(1+\delta_{\ell'_1\ell'_2})}{2L+1}}
\tilde v^L_{\ell_1\ell_2\ell'_1\ell'_2}.
\label{e_v2}
\end{equation}
Since the bosons are necessarily symmetrically coupled,
allowed two-boson states are
$s^2$ ($L=0$), $sd$ ($L=2$) and $d^2$ ($L=0,2,4$).
Since for $n$ states with a given angular momentum
one has $n(n+1)/2$ interactions,
seven independent two-body interactions $v$ are found:
three for $L=0$,
three for $L=2$
and one for $L=4$.

This analysis can be extended to higher-order interactions.
One may consider, for example, the three-body interactions
$\langle\ell_1\ell_2\ell_3;LM|\hat H^{(3)}|\ell'_1\ell'_2\ell'_3;LM\rangle$.
The allowed three-boson states are
$s^3$ ($L=0$),
$s^2d$ ($L=2$),
$sd^2$ ($L=0,2,4$)
and $d^3$ ($L=0,2,3,4,6$),
leading to $6+6+1+3+1=17$ independent three-body interactions
for $L=0,2,3,4,6$, respectively.
Note that any three-boson state $s^id^{3-i}$
is fully characterized by its angular momentum $L$;
this is no longer the case for higher boson numbers
when additional labels must be introduced.

The number of possible interactions at each order $n$
is summarized in table~\ref{t_numint} for up to $n=3$.
\begin{table}
\caption{Enumeration of $n$-body interactions in \mbox{IBM-1} for $n\leq3$.}
\label{t_numint}
\vspace{3mm}
\begin{tabular}{lcrcrcr}
\hline
Order&&\multicolumn{5}{c}{Number of interactions}\\
\cline{3-7}
&\qquad\qquad\qquad&total
&\qquad\qquad\qquad&type I$^a$
&\qquad\qquad\qquad&type II$^b$\\
\hline
$n=0$ &&  1 && 1 &&  0\\
$n=1$ &&  2 && 1 &&  1\\
$n=2$ &&  7 && 2 &&  5\\
$n=3$ && 17 && 7 && 10\\
\hline
\multicolumn{7}{l}{$^a$Interaction energy
is constant for all states with the same $N$.}\\
\multicolumn{7}{l}{$^b$Interaction energy
varies from state to state.}
\end{tabular}
\end{table}
Some of these interactions exclusively contribute to the binding energy
and do not influence the excitation spectrum of a nucleus.
To determine the number of such interactions,
one notes that the hamiltonian $\hat N\hat H^{(n-1)}$
for constant boson number ({\it i.e.}, a single nucleus)
essentially reduces to the $(n-1)$-body hamiltonian $\hat H^{(n-1)}$.
Consequently, of the $N_n$
independent interactions of order $n$ contained in $\hat H^{(n)}$,
$N_{n-1}$ terms of the type $\hat N\hat H^{(n-1)}$ must be discarded
if one wishes to retain only those that influence the excitation energies.
For example, given that there is one term of order zero
({\it i.e.}, a constant),
one of the two first-order terms
({\it i.e.}, the combination $\hat N$)
does not influence the excitation spectrum.
Likewise, there are two first-order terms
({\it i.e.}, $\hat n_s$ and $\hat n_d$)
and hence two of the seven two-body interactions
do not influence the excitation spectrum.
This argument leads to the numbers quoted in table~\ref{t_numint}.

We conclude that,
in the nucleus-by-nucleus fits that will be performed in this work,
there is a single one-boson energy of relevance,
as well as five two-body and ten three-body interactions.
This number of independent terms is too high for practical applications
and simplifications must be sought
on the basis of physical, empirical or formal arguments.
Some of them are based on the classical limit of the \mbox{IBM-1}
to which we now turn.

\subsection{The classical limit}
\label{ss_climit}
The coherent-state formalism~\cite{Ginocchio80,Dieperink80,Bohr80}
represents a bridge between algebraic and geometric nuclear models.
The central outcome of the formalism is that for any \mbox{IBM-1} hamiltonian
a corresponding potential $V(\beta,\gamma)$ can be constructed
where $\beta$ and $\gamma$ parametrize
the intrinsic quadrupole deformation of the nucleus~\cite{BM75}.
This procedure is known as the classical limit of the \mbox{IBM-1}.

The coherent states used for obtaining the classical limit of the \mbox{IBM-1}
are of the form
\begin{equation}
|N;\alpha_\mu\rangle\propto
\left(s^\dag+\sum_\mu\alpha_\mu d^\dag_\mu\right)^N|{\rm o}\rangle,
\label{e_coh}
\end{equation}
where $|{\rm o}\rangle$ is the boson vacuum
and $\alpha_\mu$ are five complex variables.
These have the interpretation of (quadrupole) shape variables
and their associated conjugate momenta.
If one limits oneself to static problems,
the $\alpha_\mu$ can be taken as real;
they specify a shape
and are analogous to the shape variables
of the droplet model of the nucleus~\cite{BM75}.
The $\alpha_\mu$ can be related to three Euler angles
which define the orientation of an intrinsic frame of reference,
and two intrinsic shape variables, $\beta$ and $\gamma$,
that parametrize quadrupole vibrations
of the nuclear surface around an equilibrium shape.
In terms of the latter variables,
the coherent state~(\ref{e_coh}) is rewritten as
\begin{equation}
|N;\beta\gamma\rangle\propto
\left(s^\dag+
\beta\left[\cos\gamma d^\dag_0
+\sqrt{\frac 1 2}\sin\gamma(d^\dag_{-2}+d^\dag_{+2})\right]
\right)^N|{\rm o}\rangle.
\label{e_cohb}
\end{equation}
The expectation value of the hamiltonian~(\ref{e_ham}) in this state
can be determined by elementary methods~\cite{Isacker81}
and yields a function of $\beta$ and $\gamma$
which is identified with a potential $V(\beta,\gamma)$,
familiar from the geometric model.
In this way the following classical limit of the hamiltonian~(\ref{e_ham}) is found:
\begin{equation}
V(\beta,\gamma)=
E_0+\sum_{n\geq1}\frac{N(N-1)\cdots(N-n+1)}{(1+\beta^2)^n}
\sum_{kl}a^{(n)}_{kl}\beta^{2k+3l}\cos^l3\gamma,
\label{e_climit}
\end{equation}
where the non-zero coefficients $a^{(n)}_{kl}$
of order $n=1$, 2 and 3 are given by
\begin{eqnarray}&&
a^{(1)}_{00}=\epsilon_s,
\quad
a^{(1)}_{10}=\epsilon_d,
\nonumber\\&&
a^{(2)}_{00}={\frac 1 2}v_{ssss}^0,
\quad
a^{(2)}_{10}=\sqrt{\frac 1 5}v_{ssdd}^0+v_{sdsd}^2,
\quad
a^{(2)}_{01}=-{\frac{2}{\sqrt 7}}v_{sddd}^2,
\nonumber\\&&
a^{(2)}_{20}=
{\frac{1}{10}}v_{dddd}^0+{\frac 1 7}v_{dddd}^2+{\frac{9}{35}}v_{dddd}^4,
\nonumber\\&&
a^{(3)}_{00}={\frac 1 6}v_{ssssss}^0,
\quad
a^{(3)}_{10}=\sqrt{\frac{1}{15}}v_{ssssdd}^0+{\frac 1 2}v_{ssdssd}^2,
\nonumber\\&&
a^{(3)}_{01}=-{\frac 1 3}\sqrt{\frac{2}{35}}v_{sssddd}^0-\sqrt{\frac 2 7}v_{ssdsdd}^2,
\nonumber\\&&
a^{(3)}_{20}=
{\frac{1}{10}}v_{sddsdd}^0+
\sqrt{\frac 1 7}v_{ssdddd}^2+
{\frac 1 7}v_{sddsdd}^2+
{\frac{9}{35}}v_{sddsdd}^4,
\nonumber\\&&
a^{(3)}_{11}=
-{\frac 1 5}\sqrt{\frac{2}{21}}v_{sddddd}^0
-{\frac{\sqrt 2}{7}}v_{sddddd}^2
-{\frac{18}{35}}\sqrt{\frac{2}{11}}v_{sddddd}^4,
\nonumber\\&&
a^{(3)}_{30}=
{\frac{1}{14}}v_{dddddd}^2
+{\frac{1}{30}}v_{dddddd}^3
+{\frac{3}{154}}v_{dddddd}^4
+{\frac{7}{165}}v_{dddddd}^6,
\nonumber\\&&
a^{(3)}_{02}=
{\frac{1}{105}}v_{dddddd}^0
-{\frac{1}{30}}v_{dddddd}^3
+{\frac{3}{110}}v_{dddddd}^4
-{\frac{4}{1155}}v_{dddddd}^6,
\label{e_coeff}
\end{eqnarray}
in terms of the single boson energies $\epsilon_s$ and $\epsilon_d$,
and the matrix elements between normalized two- and three-body states,
\begin{eqnarray}
v^L_{\ell_1\ell_2\ell'_1\ell'_2}&=&
\langle\ell_1\ell_2;LM|\hat H^{(2)}|\ell'_1\ell'_2;LM\rangle,
\nonumber\\
v^L_{\ell_1\ell_2\ell_3\ell'_1\ell'_2\ell'_3}&=&
\langle\ell_1\ell_2\ell_3;LM|\hat H^{(3)}|\ell'_1\ell'_2\ell'_3;LM\rangle.
\end{eqnarray}
The expressions~(\ref{e_climit}) and (\ref{e_coeff})
will be useful for making a choice
between the many possible three-body interactions.

\subsection{A specific two-body hamiltonian}
\label{ss_specific2}
From a great number of standard \mbox{IBM-1} studies~\cite{Iachello87}
one has a good idea of a workable hamiltonian
with up to two-body interactions which is of the form
\begin{equation}
\hat H^{(1+2)}=
\epsilon_d\,\hat n_d+
\kappa\,\hat Q\cdot\hat Q+
\kappa'\hat L\cdot\hat L+
\lambda_d\,\hat n_d^2,
\label{e_ham12}
\end{equation}
where $\hat Q$ is the quadrupole operator with components
\begin{equation}
\hat Q_\mu=
[d^\dag\times\tilde s+s^\dag\times\tilde d]^{(2)}_\mu+
\chi[d^\dag\times\tilde d]^{(2)}_\mu,
\label{e_quad}
\end{equation}
and $\hat L$ is the angular momentum operator,
$\hat L_\mu=\sqrt{10}\,[d^\dag\times\tilde d]^{(1)}_\mu$.
The $\hat Q^2$ and $\hat L^2$ terms in~(\ref{e_ham12})
constitute the hamiltonian
of the so-called consistent-$Q$ formalism (CQF)~\cite{Warner82}.
Its eigenfunctions are fully determined by $\chi$
which for $\chi=\pm\sqrt{7}/2$ gives rise to the deformed or SU(3) limit 
and for $\chi=0$ to the $\gamma$-unstable or SO(6) limit.
In an extended consistent-$Q$ formalism (ECQF)~\cite{Lipas85}
a further term $\hat n_d$ is added
with which the third, vibrational or U(5) limit of the \mbox{IBM-1}
can be obtained.
The ECQF hamiltonian thus allows one
to reach all three limits of the model with four parameters.
In some nuclei an additional term $\lambda_d\,\hat n_d^2$
further improves the description of the excitation spectrum.
The effect of this term with $\lambda_d<0$
is an increase of the moment of inertia
with increasing angular momentum (or $d$-boson seniority $\tau$).
This so-called `$\tau$-compression'
has been used for the first time in ref.~\cite{Pan92}.

For the calculation of electric quadrupole properties
an E2 transition operator is needed.
In the \mbox{IBM-1} it is defined as
$\hat T_\mu({\rm E2})=e_{\rm b}\hat Q_\mu$
where $e_{\rm b}$ is an effective charge for the bosons.
In CQF the quadrupole operator in the E2 operator
and in the hamiltonian are the same~\cite{Warner82},
that is, they contain the same $\chi$.

\subsection{A specific three-body hamiltonian}
\label{ss_specific3}
Many nuclear properties can be correctly described
by the relatively simple hamiltonian~(\ref{e_ham12})
but some cannot.
A notable example is the even--odd staggering
in the $\gamma$ band of nuclei that are close to the SO(6) limit.
A characteristic feature of the $\gamma$-unstable limit of the \mbox{IBM-1}
is a bunching of $\gamma$-band states~\cite{Arima79}
according to $2^+$, $(3^+,4^+)$,  $(5^+,6^+)$,\dots,
that is, $3^+$ and $4^+$ are close in energy, {\it etc}.
This even--odd staggering is observed in certain SO(6) nuclei but not in all
and in some it is, in fact, replaced by the opposite bunching
$(2^+,3^+)$,  $(4^+,5^+)$,\dots,
which is typical of a rigid triaxial rotor~\cite{Davydov58}.
From these qualitative observations it is clear
that the even--odd $\gamma$-band staggering
is governed by the $\gamma$ degree of freedom ({\it i.e.}, triaxiality)
as it changes character in the transition
from a $\gamma$-soft vibrator to a rigid triaxial rotor.

A proper description of triaxiality in the \mbox{IBM-1}
must necessarily involve higher-order interactions
as can be shown from the expressions given in sect.~\ref{ss_climit}.
The minimum of the potential $V(\beta,\gamma)$ in~(\ref{e_climit})
(which can be thought of as the equilibrium shape of the nucleus)
of an \mbox{IBM-1} hamiltonian with up to two-body interactions
is either spherical ($\beta=0$),
prolate deformed ($\beta>0,\gamma=0^\circ$)
or oblate deformed ($\beta>0,\gamma=60^\circ$).
The lowest term in~(\ref{e_climit})
with a triaxial extremum is quadratic in $\cos3\gamma$ ($l=2$)
and this requires a non-zero $a^{(3)}_{02}$ coefficient. 
From the explicit expressions given in eqs.~(\ref{e_coeff})
it is seen that the lowest-order interactions
possibly leading to a triaxial minimum in $V(\beta,\gamma)$
are thus necessarily of the form
\begin{equation}
\hat H_d^{(3)}=
\sum_L \tilde v_{dddddd}^L
[[d^\dag\times d^\dag]^{(\lambda)}\times d^\dag]^{(L)}
\cdot
[[\tilde d\times\tilde d]^{(\lambda')}\times\tilde d]^{(L)},
\label{e_ham3}
\end{equation}
where the allowed angular momenta are $L=0,2,3,4,6$.
For several $L$ more than one combination
of intermediate angular momenta $\lambda$ and $\lambda'$ is possible;
these do not give rise to independent terms but differ by a scale factor.
To avoid the confusion caused by this scale factor,
we rewrite the hamiltonian~(\ref{e_ham3}) as
\begin{equation}
\hat H_d^{(3)}=
\sum_L v_{dddddd}^L\hat B_L^\dag\cdot\tilde B_L,
\qquad
\hat B_{LM}^\dag=N_{\lambda L}
[[d^\dag\times d^\dag]^{(\lambda)}\times d^\dag]^{(L)}_M.
\label{e_ham3b}
\end{equation}
For simplicity's sake the coefficients $v_{dddddd}^L$
shall be denoted as $v_L$ in the following.
The normalization coefficient $N_{\lambda L}$ is defined such
that $B_{LM}|d^3;LM\rangle$ yields the vacuum state $|{\rm o}\rangle$,
where $|d^3;LM\rangle$ is a normalized, symmetric state
of three bosons coupled to total angular momentum $L$
and $z$ projection $M$.
The normalization coefficients $N_{\lambda L}$
are given in table~\ref{t_norm} for the different combinations
of $\lambda$ and $L$.
\begin{table}
\caption{
Normalization coefficients $N_{\lambda L}$
for three-$d$-boson states.}
\vspace{3mm}
\label{t_norm}
\begin{tabular}{ccccccccccc}
\hline
$L$&\qquad&0&~~&2&~~&3&~~&4&~~&6\\
\hline
$\lambda=0$
&&---&&$\sqrt{\frac{5}{14}}$&&---&&---&&---\\
$\lambda=2$
&&$\sqrt{\frac{1}{6}}$&&$\sqrt{\frac{7}{8}}$
&&$\phantom{-}\sqrt{\frac{7}{30}}$&&$\sqrt{\frac{7}{22}}$&&---\\
$\lambda=4$
&&---&&$\sqrt{\frac{35}{72}}$&&$-\sqrt{\frac{7}{12}}$
&&$\sqrt{\frac{7}{20}}$&&$\sqrt{\frac{1}{6}}$\\
\hline
\end{tabular}
\end{table}
Results are independent of $\lambda$
provided the appropriate coefficient $N_{\lambda L}$ is used.

While there are good arguments
for choosing any of the three-body terms $\hat B_L^\dag\cdot\tilde B_L$,
it is more difficult to distinguish {\it a priori}
between these five different interactions.
From the expression for $a^{(3)}_{02}$ given in eqs.~(\ref{e_coeff})
it is seen that the cubic term $\hat B_L^\dag\cdot\tilde B_L$ with $L=3$
is proportional to $\sin^23\gamma$.
It is therefore the interaction which is most effective
to create a triaxial minimum in the potential $V(\beta,\gamma)$
and for this reason it has been studied in most detail.
The effect of $\hat B_3^\dag\cdot\tilde B_3$
on even--odd staggering in the $\gamma$ band
was demonstrated with numerical calculations~\cite{Heyde84}.
Applications of the $L=3$ three-body term
were proposed in ref.~\cite{Casten85b} for SO(6)-like Xe and Ba isotopes
in the mass region around $A=130$, as well as for $^{196}$Pt.

Besides these physical and empirical arguments,
there are also attractive formal aspects
of the $\hat B_3^\dag\cdot\tilde B_3$ interaction among the $d$ bosons.
A first one concerns its effect in the SU(3) limit of the \mbox{IBM-1}.
In this limit states are characterized by the U(6) label $[N]$,
the SU(3) labels $(\lambda,\mu)$,
the angular momentum or SO(3) label $L$ and its $z$ projection $M$,
and by an ${\rm SU}(3)\supset{\rm SO}(3)$
multiplicity label $\kappa$~\cite{Arima78}.
Likewise, any interaction can be written
in terms of products of tensor operators
$\hat T^\dag_{[N](\lambda,\mu)\kappa LM}$
[which creates an $N$-boson state
with the quantum numbers $(\lambda,\mu)\kappa LM$]
and their hermitian conjugates.
Since a three-boson state with $L=3$ is unique,
it follows that its SU(3) labels are fixed, namely,
$(\lambda,\mu)=(2,2)$ and $\kappa=2$,
and that the following proportionality must hold:
\begin{equation}
B^\dag_{L=3,M}\propto
\hat T^\dag_{[3](2,2)\kappa=2,L=3,M}.
\label{e_v3}
\end{equation}
This property can be used to show that
\begin{eqnarray}
&&\tilde B_{L=3,M}|[N](2N,0)L'M'\rangle=0,
\nonumber\\
&&\tilde B_{L=3,M}|[N](2N-4,2)\kappa'=0,L'M'\rangle=0,
\label{e_v3su3}
\end{eqnarray}
that is, the $\hat B_3^\dag\cdot\tilde B_3$ interaction
acting on the ground-state band $(2N,0)$
or on the $\beta$-vibrational band $(2N-4,2)\kappa=0$ gives zero.
The former property results from the fact
that the SU(3) Kronecker product $(2N,0)\times(2,2)$ does not yield
an SU(3) representation that is contained in the U(6) representation $[N-3]$.

The situation can be summarized by stating that
the SU(3) hamiltonian augmented
with a $\hat B_3^\dag\cdot\tilde B_3$ interaction
is an example of a partial dynamical symmetry~\cite{Alhassid92}:
while the eigenstates of this extended hamiltonian
are not solvable in general, some of them are,
in particular the members of the ground-state and $\beta$ bands.
In first approximation the effect of $\hat B_3^\dag\cdot\tilde B_3$
is to shift the entire $\gamma$ band in energy
without changing its moment of inertia or the structure of its states.
We also note that this nicely complements
the (two-body) interaction derived previously~\cite{Leviatan96}
which leaves the ground-state and $\gamma$ bands solvable
but modifies the structure of the $\beta$ band.

\begin{figure*}
\vspace*{5mm}
\begin{center}
\includegraphics[width=9cm]{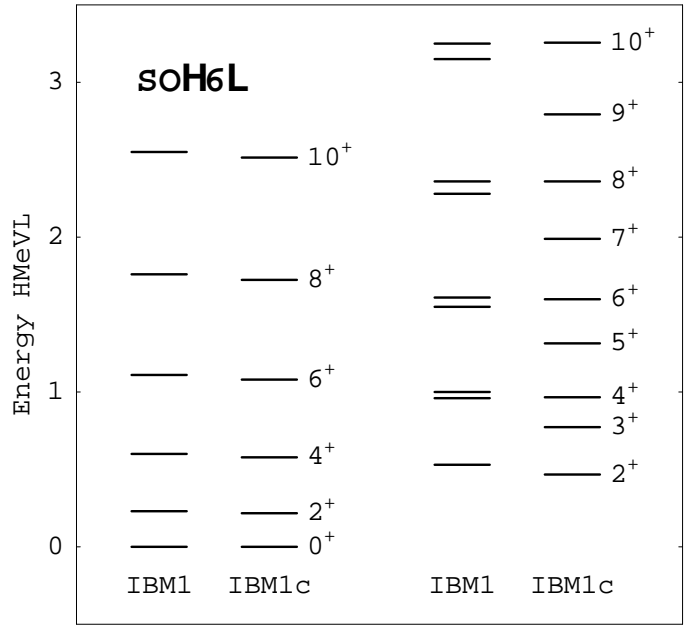}
\caption{
Levels of the ground-state and $\gamma$ bands
up to angular momentum $J^\pi=10^+$
in the exact SO(6) limit (labelled `IBM1')
and in the SO(6) limit plus the three-body interaction $\hat B_3^\dag\cdot\tilde B_3$
(labelled `IBM1c').
The \mbox{IBM-1} hamiltonian~(\ref{e_ham12}) is used for $\chi=0$
with $\kappa=-50$~keV, $\kappa'=5$~keV and $\epsilon_d=\lambda_d=0$;
the strength of the three-body term is $v_3=-50$~keV.
The number of bosons is $N=10$.}
\label{f_so6e}
\end{center}
\end{figure*}

\begin{figure*}
\vspace*{5mm}
\begin{center}
\includegraphics[width=9cm]{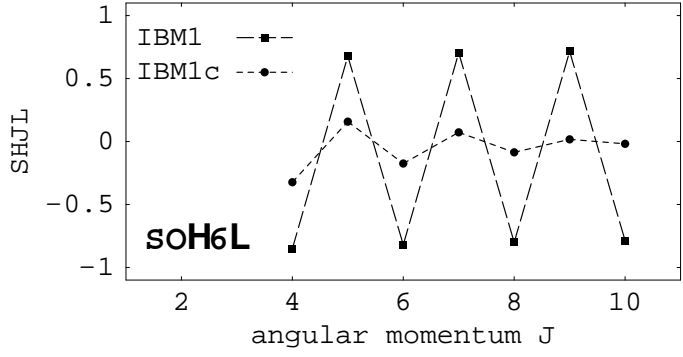}
\caption{
The signature splitting
of the $\gamma$ band in the exact SO(6) limit (labelled `IBM1')
and in the SO(6) limit plus the three-body interaction $\hat B_3^\dag\cdot\tilde B_3$
(labelled `IBM1c').
The parameters of the hamiltonian
are given in the caption of fig.~\ref{f_so6e}.}
\label{f_so6s}
\end{center}
\end{figure*}

\begin{figure*}
\vspace*{5mm}
\begin{center}
\includegraphics[width=6.5cm]{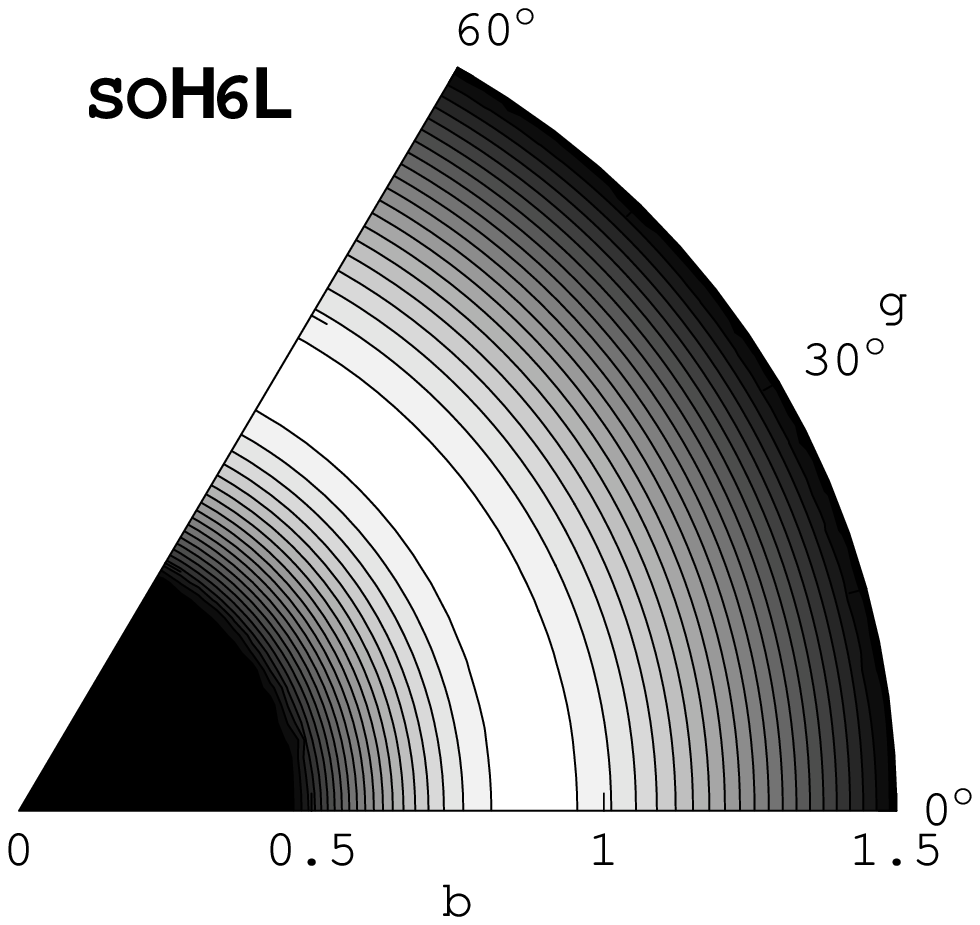}
\includegraphics[width=6.5cm]{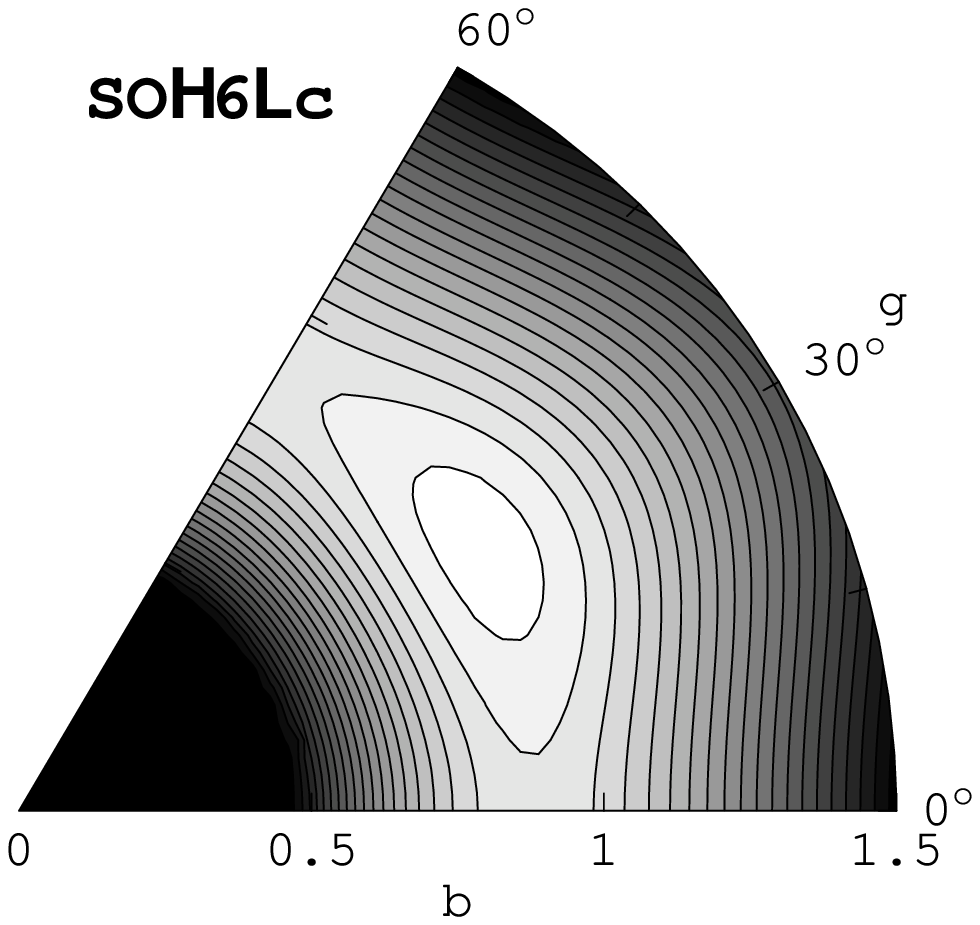}
\caption{
Potential energy surfaces $V(\beta,\gamma)$ in the SO(6) limit.
The plot on the left-hand side
shows the classical limit of the \mbox{IBM-1} hamiltonian in the SO(6) limit
while on the right-hand side
the effect of $\hat B_3^\dag\cdot\tilde B_3$ is included.
The parameters of the hamiltonian
are given in the caption of fig.~\ref{f_so6e}.}
\label{f_so6p}
\end{center}
\end{figure*}

Let us now turn to the effect
of the $\hat B_3^\dag\cdot\tilde B_3$ interaction
in the SO(6) limit of the \mbox{IBM-1}.
Its influence on the energy spectrum
is illustrated in fig.~\ref{f_so6e}.
It is seen that the ground-state band levels
are only slightly affected by the cubic interaction
while the effect on the $\gamma$-band energies is important.
In particular, the even--odd staggering in the $\gamma$ band,
characteristic of $\gamma$-soft or SO(6) behaviour,
is greatly diminished.
A sensitive way of testing the signature splitting of the $\gamma$ band
is through $S(J)$ given by~\cite{Zamfir91}
\begin{equation}
S(J)=
\frac{E(J)-E(J-1)}{E(J)-E(J-2)}\cdot
\frac{J(J+1)-(J-1)(J-2)}{J(J+1)-J(J-1)}-1,
\label{e_stag}
\end{equation}
which vanishes if there is no even--odd staggering.
This quantity is shown in fig.~\ref{f_so6s}
for the schematic SO(6) case.
The figure confirms that the three-body interaction
$\hat B_3^\dag\cdot\tilde B_3$ (with $v_3<0$)
has the property of reducing the $\gamma$-band staggering
but it also reveals that this reduction is more important
for the high-$J$ levels.
This is a characteristic feature
that can be experimentally tested.
Figure~\ref{f_so6p} illustrates the effect
of the cubic interaction $\hat B_3^\dag\cdot\tilde B_3$
on the potential $V(\beta,\gamma)$ derived in the classical limit.
While the potential in the SO(6) limit is completely $\gamma$ independent,
a shallow triaxial minimum develops for $v_3<0$.
We note, however, that this triaxial minimum
exists precariously for $\chi=0$
and quickly disappears when $\chi$ acquires non-zero values.

Most of the results presented below
are obtained with the $d$-boson cubic interaction with $L=3$
which in general reproduces best the $\gamma$-band properties.
We have nevertheless systematically investigated the terms with $L\neq3$,
and will occasionally refer to those results in the following.

\section{Numerical procedure}
\label{s_num}
To test effectiveness of the various cubic interactions
in reproducing the data in near-SO(6) nuclei,
we have devised the following fitting procedure.
The nuclei considered should have enough known states
in the ground-state and $\gamma$ bands---preferably
up to angular momentum $J^\pi=10^+$---for
the procedure to be meaningful.
The first step is to determine the parameters
in the standard \mbox{IBM-1} hamiltonian~(\ref{e_ham12}).
For an initial choice of $\chi$,
the parameters $\kappa$ and $\kappa'$ are first determined
while keeping $\epsilon_d$ and $\lambda_d$ zero.
With ($\kappa,\kappa'$) thus found as starting values,
a new fit is performed setting $\epsilon_d$ free as well,
leading to the best values ($\kappa,\kappa',\epsilon_d$).
Finally, this process is repeated by letting also $\lambda_d$ free,
leading to a final set ($\kappa,\kappa',\epsilon_d,\lambda_d$) for a given $\chi$.
In some nuclei, especially if not enough data are available,
the last step can prove numerically unstable
and no unique set ($\kappa,\kappa',\epsilon_d,\lambda_d$) is found.
In that case we leave $\lambda_d$ equal to zero.
The parameter $\chi$ cannot be reliably determined from energies
but is fixed from E2 transition rates which are calculated in CQF.
If not enough E2 data are available,
we take $\chi$ from a neighbouring isotope.
The entire procedure is repeated for different $\chi$,
retaining the value that gives the best agreement with the E2 data.
In a last step the importance of the $\hat B_L^\dag\cdot\tilde B_L$ terms
is tested in a similar way by allowing the variation of all five parameters
($\kappa,\kappa',\epsilon_d,\lambda_d,v_L$)---or
four if $\lambda_d=0$---while keeping $\chi$ constant.
Since we are particularly interested
in the influence of $v_L$ on the even--odd staggering,
in this final step we adjust this parameter
while assigning a larger weight
to the members of the $\gamma$ band.
The accuracy of the fits can be tested
by plotting the signature splitting $S(J)$
which will be done systematically.

\section{Results and discussion}
\label{s_res}
In this section we present the results for 16 different nuclei
ranging from neutron-rich Ru and Pd isotopes,
via Xe and Ba nuclei with mass number $A\sim130$
to neutron-deficient Os and Pt isotopes.
All nuclei are close to the SO(6) limit of the \mbox{IBM-1}
and have $\gamma$ bands that are known at least up to $J^\pi=7^+$.
Since the \mbox{IBM-1} is a low-spin model,
states up to $J^\pi=10^+$ but not higher are included in the fits.
All data have been retrieved
from the Brookhaven National Nuclear Data Center~\cite{NNDC}
unless indicated otherwise.

\subsection{Ruthenium and palladium isotopes}
\label{ss_rupd}
In the context of the interacting boson model,
the Ru and Pd isotopes are described as transitional
between vibrational [or U(5)] and $\gamma$ soft [or SO(6)]~\cite{Stachel82}.
In this interpretation the neutron-rich members of these isotopic chains
are close to the SO(6) limit of the \mbox{IBM-1}
and thus they fall into the class of nuclei we wish to study in this work.

Recently, gamma-ray spectroscopy of the fission fragments
produced by a $^{252}$Cf source
has significantly improved our knowledge
of the structure of the isotopes $^{108-112}$Ru~\cite{Gore05}.
The new data that became available on these nuclei
was considered in ref.~\cite{Stefanescu07}
where particular attention was paid
to the staggering pattern in the $\gamma$ band
and the triaxial degree of freedom.
The advantage of the method as described in sect.~\ref{s_num} is
that a consistent one- and two-body \mbox{IBM-1} hamiltonian is taken
to which a three-body term is added without changing the value of $\chi$.
In this way any improvement of the description
of the $\gamma$-band staggering
can be unambiguously attributed to the three-body term.
Also, a least-squares fit is performed to the parameters
in the hamiltonian according to the procedure outlined in sect.~\ref{s_num}.
In spite of these differences
the results obtained here are globally in agreement
with those of Stefanescu {\it et al.}~\cite{Stefanescu07}.
The main conclusion is that,
while the staggering pattern of the $\gamma$ band
is much improved
with the $\hat B_3^\dag\cdot\tilde B_3$ interaction
in $^{110}$Ru and $^{112}$Ru,
this is not the case for $^{108}$Ru.
This is evident from the parameters shown in table~\ref{t_parrupd}.
The root-mean-square (rms) deviation $\sigma$ actually increases for $^{108}$Ru
when the three-body interaction is added to the hamiltonian
as a consequence of the fit procedure
which gives more weight to the $\gamma$-band members
when also $v_3$ is fitted.
\begin{table}
\caption{
Parameters and rms deviation for Ru and Pd isotopes
in units of keV.}
\label{t_parrupd}
\vspace{3mm}
\begin{tabular}{ccccccccccccccc}
\hline
Nucleus&~&
$\epsilon_d$&~&$\kappa$&~&$\kappa'$&~&$\lambda_d$&~&$v_3$&~&$\chi^*$
&~&$\sigma$\\
\hline
$^{108}$Ru
&&$1078$&&$-57.6$&&$12.1$&&$-144.9$&&---&&$-0.10$&&$23$\\
&&$~852$&&$-66.8$&&$~8.3$&&$-130.7$&&$-13.1$&&$-0.10$&&$45$\\
&&$~732$&&$-74.6$&&$14.0$&&$-157.8$&&$~~~~~30.5^{**}$&&$-0.10$&&$19$\\
$^{110}$Ru
&&$1053$&&$-46.1$&&$15.5$&&$-123.7$&&---&&$-0.10$&&$39$\\
&&$~873$&&$-56.9$&&$~9.9$&&$-108.5$&&$-28.1$&&$-0.10$&&$20$\\
$^{112}$Ru
&&$~837$&&$-45.3$&&$15.2$&&$-116.8$&&---&&$-0.10$&&$55$\\
&&$~424$&&$-57.8$&&$~7.7$&&~$-73.7$&&$-46.8$&&$-0.10$&&$38$\\
\hline
$^{114}$Pd
&&$~321$&&$~~~1.0$&&$12.2$&&---&&---&&$-0.10$&&$91$\\
&&$~486$&&$-19.6$&&$~5.7$&&---&&$-94.6$&&$-0.10$&&$61$\\
$^{116}$Pd
&&$~367$&&~$-3.5$&&$12.4$&&---&&---&&$-0.10$&&$85$\\
&&$~451$&&$-19.1$&&$~8.3$&&---&&$-84.3$&&$-0.10$&&$56$\\
\hline
\multicolumn{15}{l}{$^*$Dimensionless. $^{**}$Value of the coefficient $v_2$.}
\end{tabular}
\end{table}
The increase in $\sigma$ illustrates
that the $\gamma$-band energies in $^{108}$Ru cannot be reproduced
by adding $\hat B_3^\dag\cdot\tilde B_3$
without destroying the agreement for the ground-state band.

\begin{figure*}
\vspace*{5mm}
\begin{center}
\includegraphics[width=9cm]{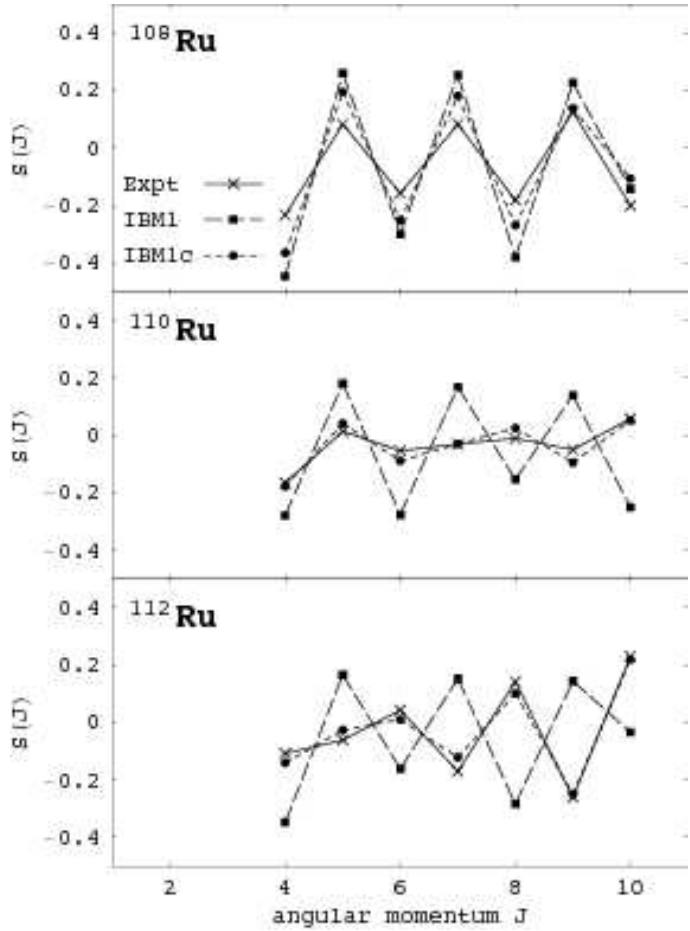}
\caption{
Observed and calculated signature splitting
of the $\gamma$ bands in $^{108-112}$Ru.
The data are indicated by crosses
and the results of the \mbox{IBM-1}
without and with the three-body interaction $\hat B_3^\dag\cdot\tilde B_3$
by squares and dots, respectively.}
\label{f_rus}
\end{center}
\end{figure*}

As one goes to the heavier Ru isotopes,
one notices a distinct evolution of the even--odd staggering pattern
(see fig.~\ref{f_rus}).
Whereas the staggering pattern is essentially consistent with the \mbox{IBM-1} calculation
without cubic interactions in $^{108}$Ru,
this is no longer the case in the two heavier isotopes.
In $^{110}$Ru there is very little staggering at all, $S(J)\approx0$,
and in $^{112}$Ru the staggering pattern in the data
is in fact the reverse of what is obtained without cubic interactions,
especially at higher angular momenta.
The $\hat B_3^\dag\cdot\tilde B_3$ interaction
shifts levels with even (odd) angular momentum upwards (downwards) in energy
and it does so increasingly with increasing spin.
This is exactly what can be observed from the data in $^{110}$Ru and $^{112}$Ru
and this provides a strong phenomenological argument
for the use of the $\hat B_3^\dag\cdot\tilde B_3$ interaction.

\begin{figure}
\centering
\includegraphics[width=9cm]{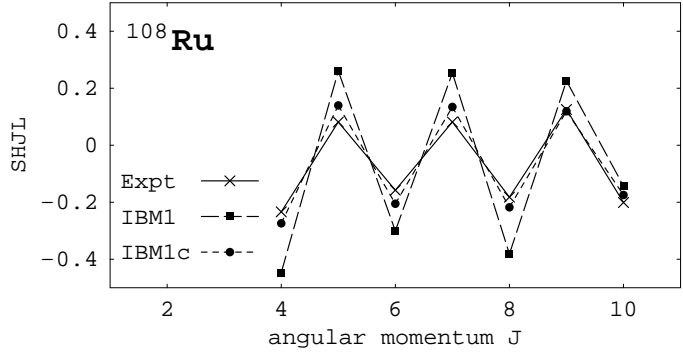}
\caption{
Same caption as fig.~\ref{f_rus} for the nucleus $^{108}$Ru
and the three-body interaction $\hat B_2^\dag\cdot\tilde B_2$.}
\label{f_ru108s2}
\end{figure}

From the plot of the signature splitting we can also `understand'
why the $\hat B_3^\dag\cdot\tilde B_3$ interaction fails in $^{108}$Ru:
the deviations in staggering between the data
and the \mbox{IBM-1} calculation without cubic interactions
actually decrease rather than increase with angular momentum.
This feature is incompatible with the $L=3$ term in the hamiltonian~(\ref{e_ham3})
but is exactly what is obtained with the $L=2$ term as shown in fig.~\ref{f_ru108s2}.

\begin{figure}
\centering
\includegraphics[width=9cm]{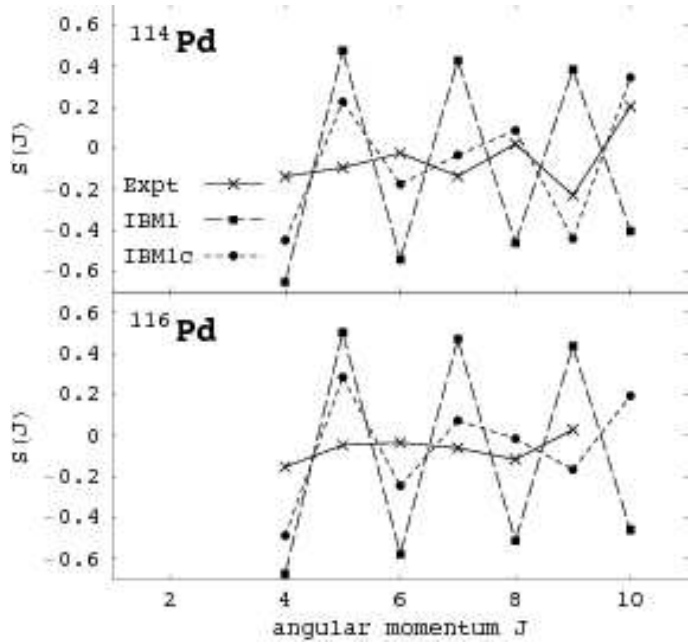}
\caption{
Same caption as fig.~\ref{f_rus} for the nuclei $^{114,116}$Pd.}
\label{f_pds}
\end{figure}

The results for the $\gamma$-band staggering in the $^{114,116}$Pd isotopes
are shown in fig.~\ref{f_pds}.
In these nuclei one cannot reliably
determine the coefficient $\lambda_d$ of $\hat n_d^2$
which is therefore kept zero.
The inclusion of the three-body interaction $\hat B_3^\dag\cdot\tilde B_3$
does improve the fit substantially
(see the rms deviations in table~\ref{t_parrupd})
but even with this term the observed staggering pattern cannot be reproduced.

It is important to check that the cubic hamiltonian thus obtained
gives reasonable results as regards electric quadrupole transitions.
This will be illustrated for some of the isotopes
discussed in the subsequent subsections.
For the Ru isotopes it has been shown to be the case in ref.~\cite{Stefanescu07}
while the relevant E2 data are not known for $^{114,116}$Pd.

\begin{figure}
\centering
\includegraphics[width=6.5cm]{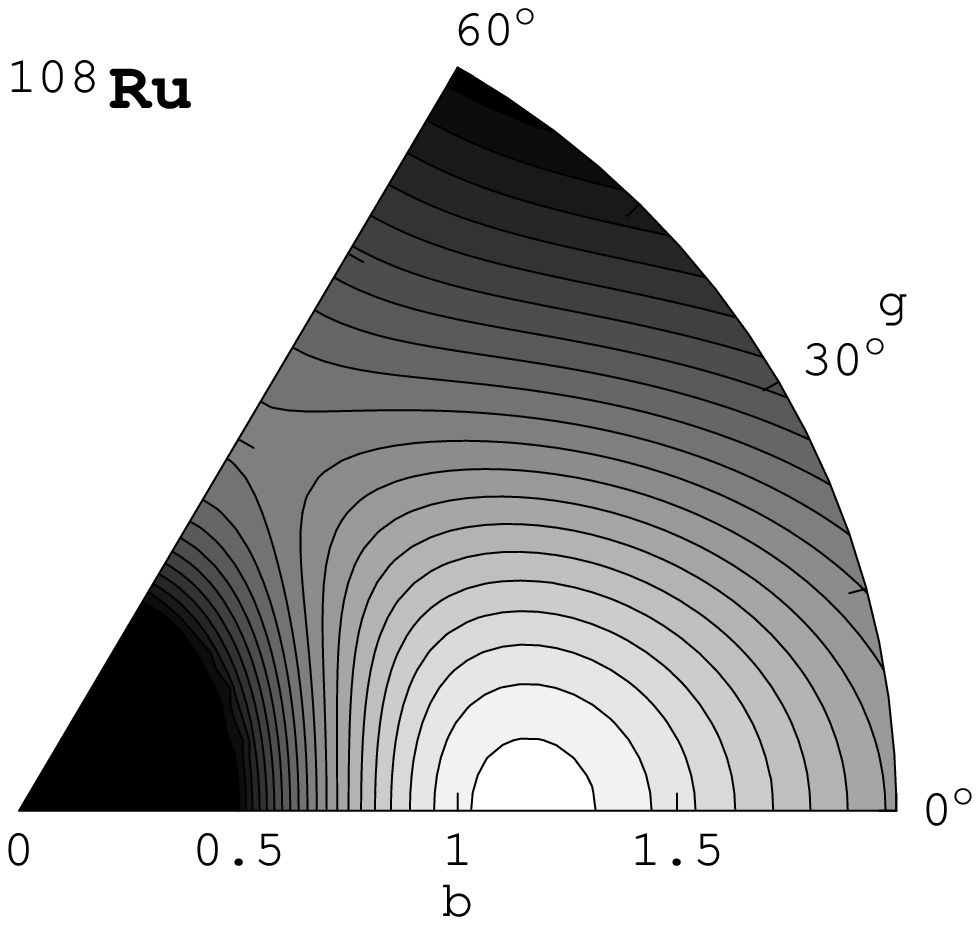}
\includegraphics[width=6.5cm]{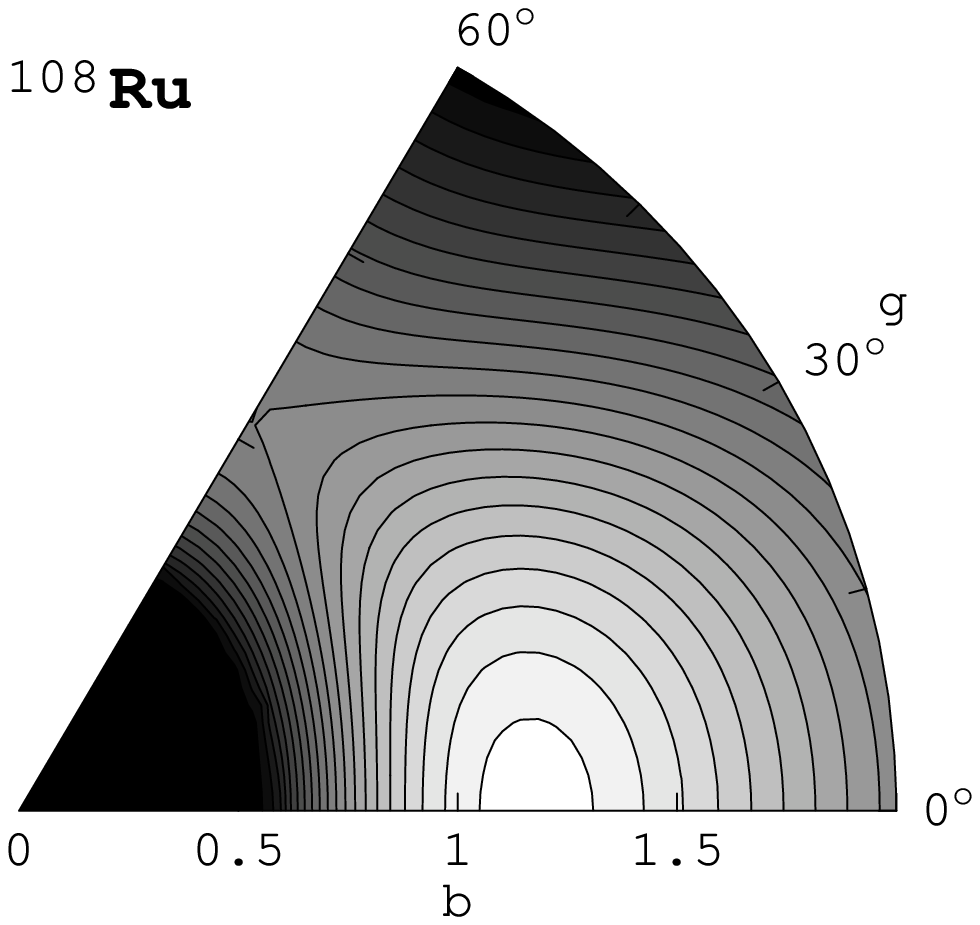}
\includegraphics[width=6.5cm]{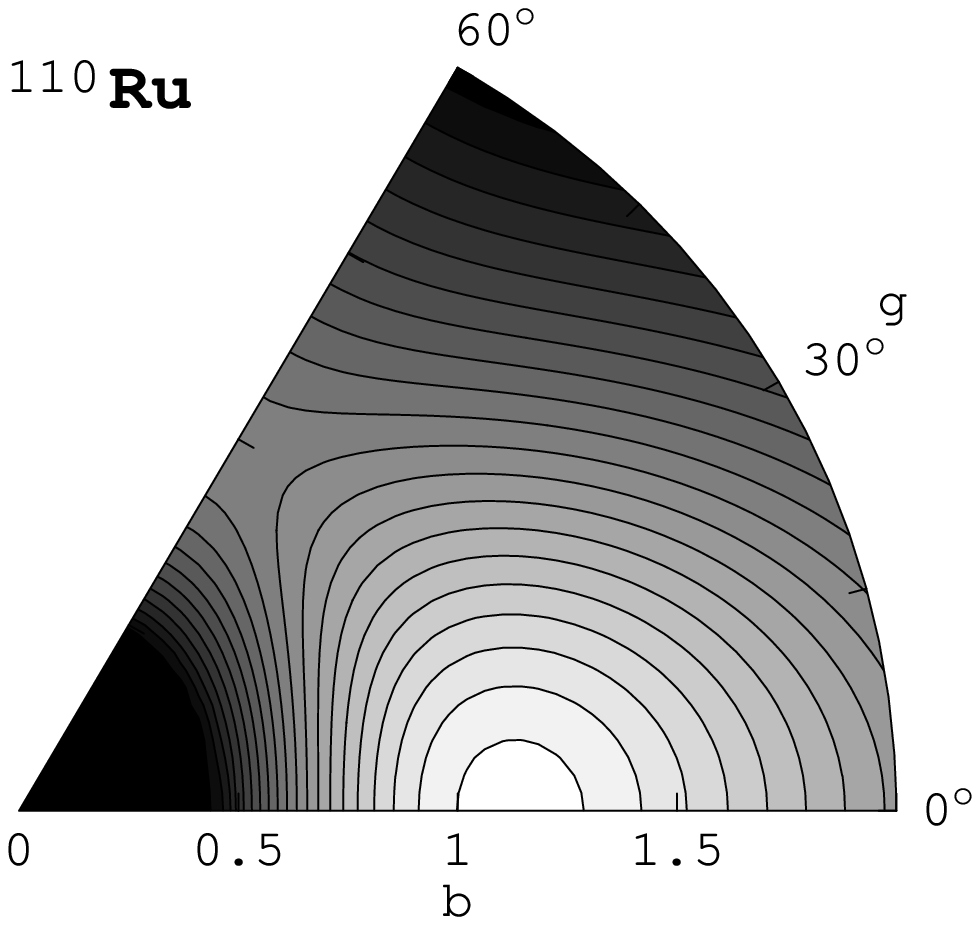}
\includegraphics[width=6.5cm]{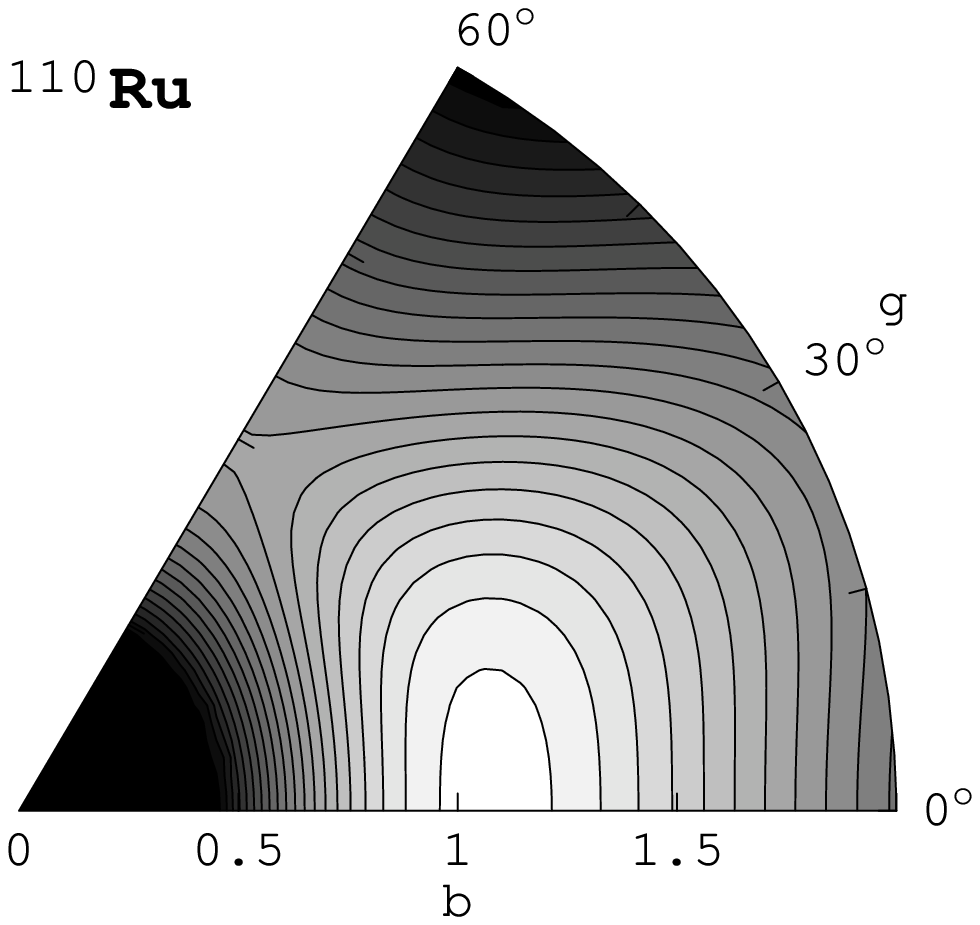}
\includegraphics[width=6.5cm]{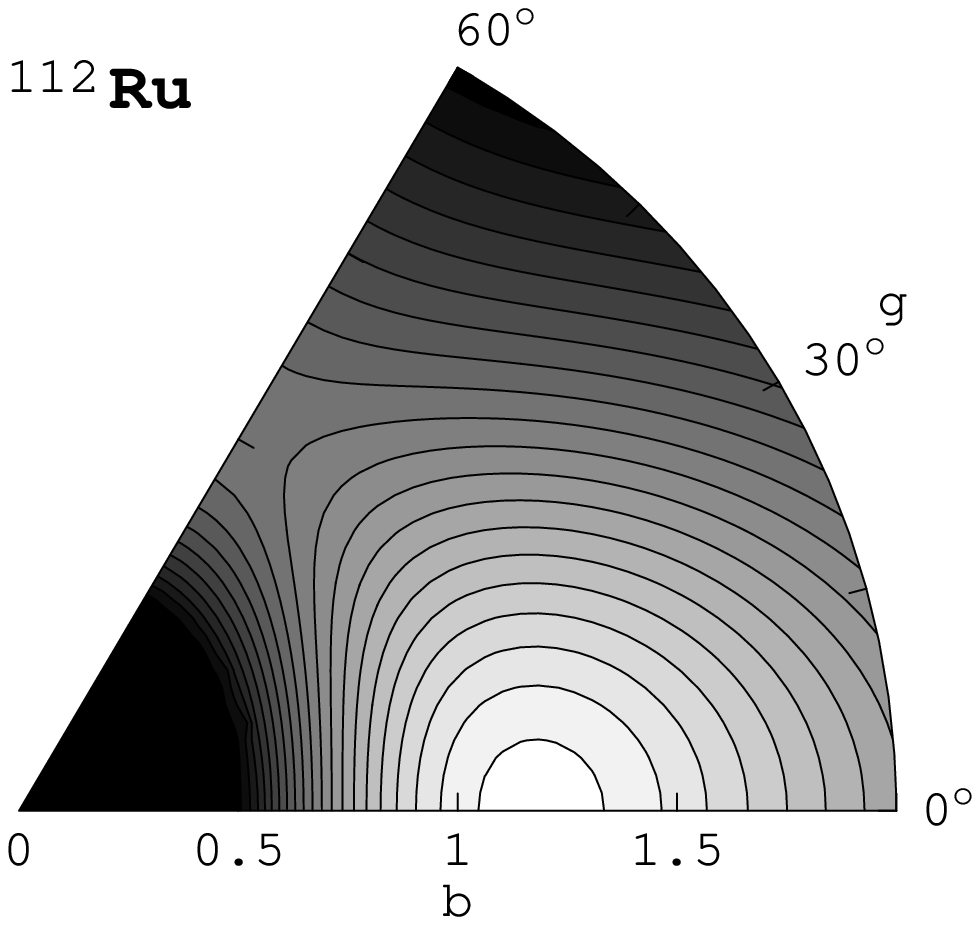}
\includegraphics[width=6.5cm]{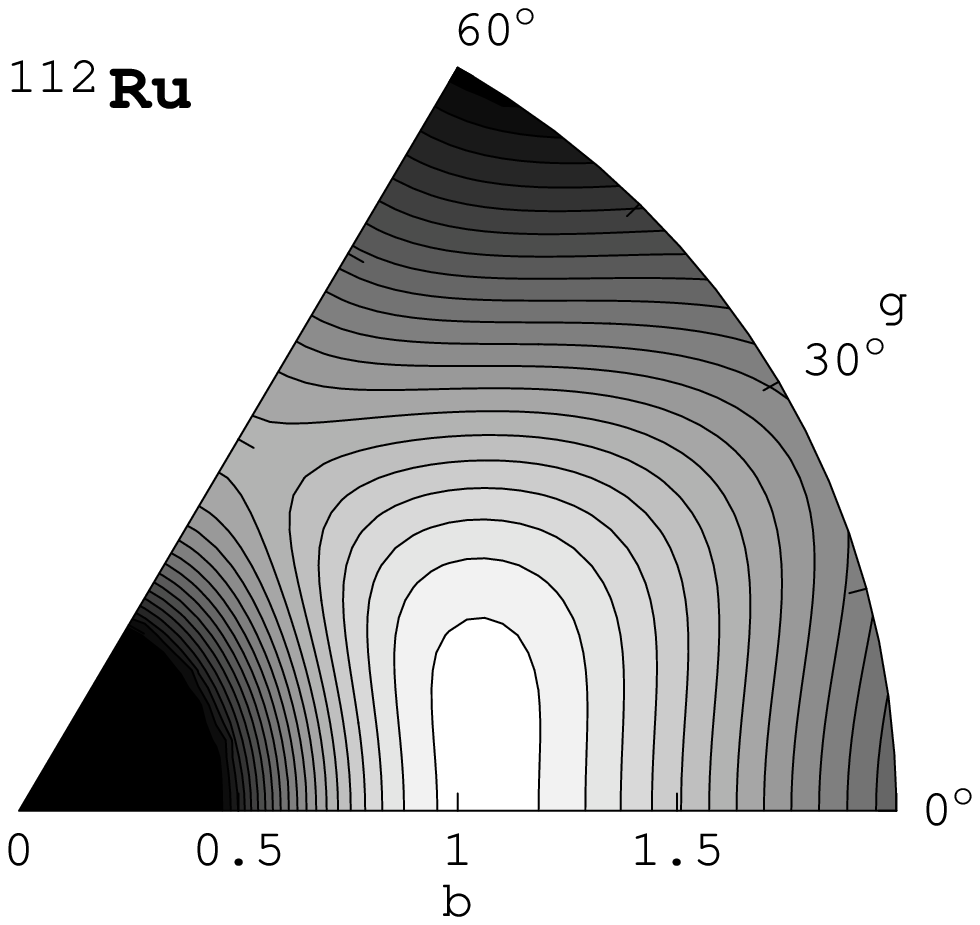}
\caption{
Potential energy surfaces $V(\beta,\gamma)$ for the Ru isotopes.
The plots on the left-hand side
show the classical limit of the \mbox{IBM-1} hamiltonian
with only two-body interactions
while on the right-hand side
the effect of $\hat B_3^\dag\cdot\tilde B_3$ is included.}
\label{f_rup}
\end{figure}

Once the parameters of the hamiltonian have been fitted
to the energy spectrum and E2 transition rates,
its classical limit yields a potential energy surface $V(\beta,\gamma)$
as obtained from the expression~(\ref{e_climit}).
In this way it can be verified to what extent triaxial features
are introduced by the cubic interactions.
Figure~\ref{f_rup} provides an illustration
by showing the potential energy surfaces $V(\beta,\gamma)$ for the Ru isotopes
obtained in the classical limit of the \mbox{IBM-1} hamiltonian
without and with the $\hat B_3^\dag\cdot\tilde B_3$ interaction.
The surfaces on the left-hand side are obtained from the two-body hamiltonian
and have a prolate minimum ($\gamma=0^{\rm o}$) for $\beta\approx1.2$.
(The quadrupole deformation parameter $\beta$ is larger in the IBM
than in the geometric Bohr--Mottelson model
for reasons discussed in ref.~\cite{Ginocchio80b}.)
The hamiltonian which includes the $\hat B_3^\dag\cdot\tilde B_3$ interaction
yields the surfaces on the right-hand side.
One notices that in the heavier Ru isotopes 
the deformed minimum extends further towards triaxial shapes
and becomes very flat in $^{112}$Ru with $\beta\approx1$ up to $\gamma\approx15^{\rm o}$.
In $^{112}$Ru noticeable changes of the potential $V(\beta,\gamma)$
are found as a result of the inclusion of cubic interactions
which perhaps is not surprising
since parameter variations are rather important
between \mbox{IBM-1} and \mbox{IBM-1c} in this nucleus (see table~\ref{t_parrupd}).
However, even in this example with a large $v_3$ parameter
no triaxial minimum is obtained.
A similar analysis of $^{114,116}$Pd yields potentials with a spherical minimum
which become flatter if the $\hat B_3^\dag\cdot\tilde B_3$ interaction
is added to the hamiltonian.

\subsection{Xenon and barium isotopes}
\label{ss_xeba}
Casten and von Brentano~\cite{Casten85a} pointed out the occurrence
of SO(6)-like nuclei in the region with mass number $A\sim130$.
In particular, the isotopes $^{124-130}$Xe and $^{128-134}$Ba
were found to display properties that are consistent with the SO(6) limit.
In a subsequent study~\cite{Casten85b} it was shown that
the theoretical description of these nuclei is greatly improved
via the inclusion of the three-body interaction $\hat B_3^\dag\cdot\tilde B_3$.
Since these early studies more members of the $\gamma$ band
have been established experimentally in several Xe and Ba isotopes
and it appears therefore worthwhile to reconsider this region
as regards triaxiality features and $\gamma$-band staggering.

\begin{figure}
\centering
\includegraphics[width=9cm]{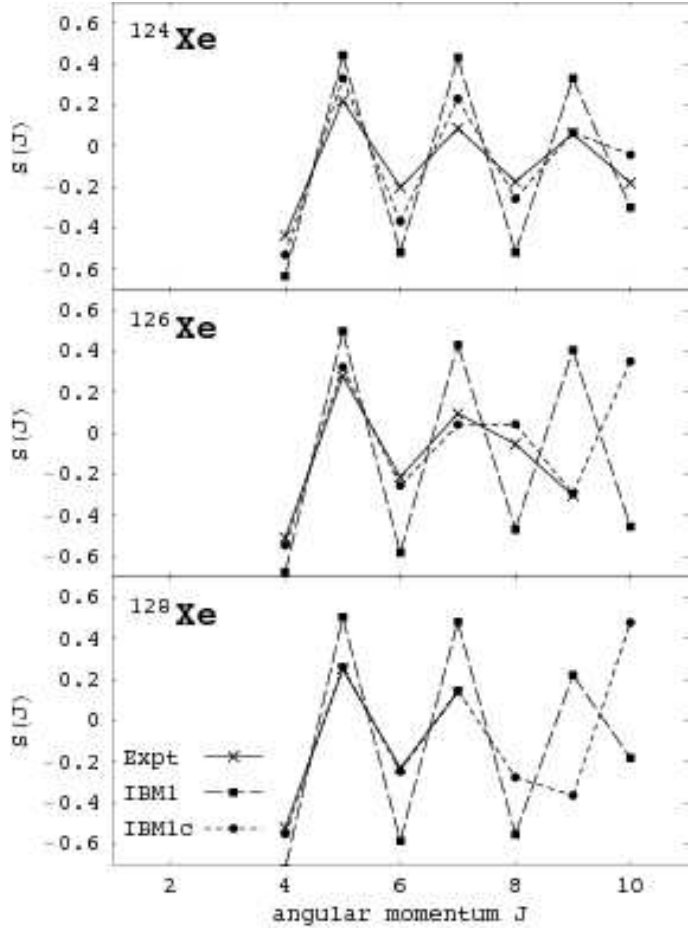}
\caption{
Same caption as fig.~\ref{f_rus} for the nuclei $^{124-128}$Xe.}
\label{f_xes}
\end{figure}

\begin{figure}
\centering
\includegraphics[width=9cm]{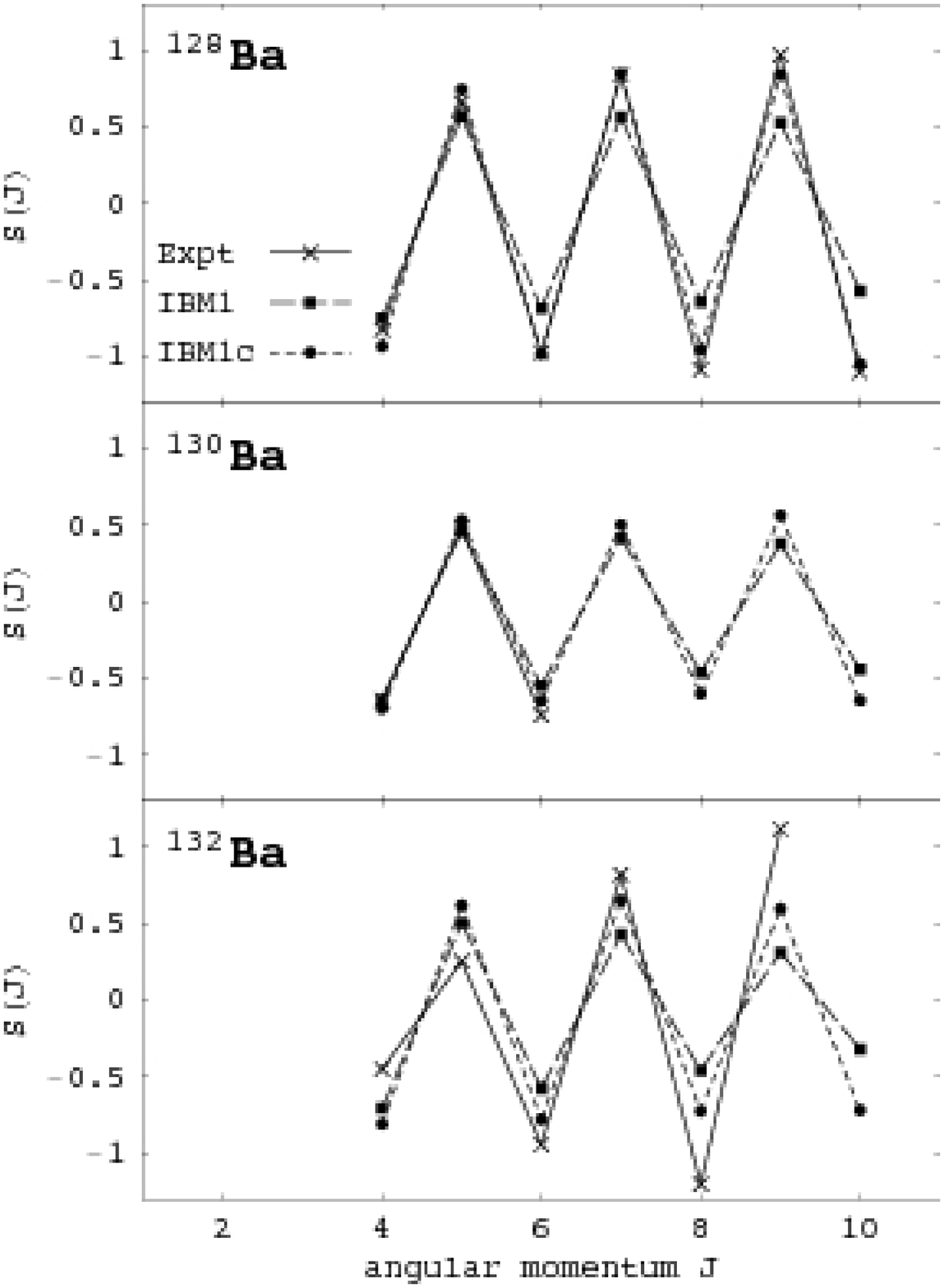}
\caption{
Same caption as fig.~\ref{f_rus} for the nuclei $^{128-132}$Ba.}
\label{f_bas}
\end{figure}

The fitting procedure as explained in sect.~\ref{s_num}
leads to the parameters and rms deviations shown in table~\ref{t_parxeba}.
\begin{table}
\caption{
Parameters and rms deviation for Xe and Ba isotopes in units of keV.}
\label{t_parxeba}
\vspace{3mm}
\begin{tabular}{ccccccccccccccc}
\hline
Nucleus&~&
$\epsilon_d$&~&$\kappa$&~&$\kappa'$&~&$\lambda_d$&~&$v_3$&~&$\chi^*$
&~&$\sigma$\\
\hline
$^{124}$Xe
&&$922$&&$-65.8$&&$11.3$&&$-145.8$&&---&&$-0.10$&&$44$\\
&&$841$&&$-59.1$&&$~8.0$&&$-111.5$&&~$-33.4$&&$-0.10$&&$24$\\
$^{126}$Xe
&&$788$&&$-49.2$&&$12.3$&&$-111.7$&&---&&$-0.10$&&$82$\\
&&$406$&&~$-7.4$&&$~6.8$&&$~~~22.7$&&$-105.1$&&$-0.10$&&$18$\\
$^{128}$Xe
&&$795$&&$-54.0$&&$11.2$&&$-130.0$&&---&&$-0.10$&&$42$\\
&&$788$&&$-69.9$&&$~8.3$&&$-141.5$&&~$-55.5$&&$-0.10$&&$14$\\
\hline
$^{128}$Ba
&&$808$&&$-55.8$&&$~7.2$&&$-107.7$&&---&&$-0.10$&&$77$\\
&&$888$&&$-62.1$&&$11.4$&&$-144.2$&&$~~~36.4$&&$-0.10$&&$50$\\
$^{130}$Ba
&&$730$&&$-48.3$&&$11.0$&&~$-98.6$&&---&&$-0.20$&&$26$\\
&&$828$&&$-51.2$&&$14.4$&&$-130.6$&&$~~~31.1$&&$-0.20$&&$20$\\
$^{132}$Ba
&&$700$&&$-23.2$&&$12.6$&&~$-70.4$&&---&&$-0.20$&&$86$\\
&&$759$&&$-26.5$&&$14.8$&&~$-94.5$&&$~~~41.5$&&$-0.20$&&$76$\\
\hline
\multicolumn{15}{l}{$^*$Dimensionless.}
\end{tabular}
\end{table}
One notes that in all Xe and Ba isotopes
the inclusion of the $\hat B_3^\dag\cdot\tilde B_3$ interaction
leads to a smaller rms deviation.
Parameter fluctuations can be large, however.
This is particularly the case in the Xe isotopes,
with parameters in $^{126}$Xe very different from those in the neighbouring isotopes.
A possible reason is the fact that the parameters in the hamiltonian
are strongly correlated
and hence very sensitive to slight changes in the fitted data.
The results for the $\gamma$-band staggering
are shown in figs.~\ref{f_xes} and~\ref{f_bas}.
In the Xe isotopes one observes an evolution
of the $\gamma$-band signature splitting with neutron number
similar to the one in the Ru isotopes.
The observed even--odd staggering is smaller
than the one calculated without cubic interaction,
and this difference increases with angular momentum $J$
to the extent that in the heavier Xe isotopes it is reversed at $J=10$.
Note that the $\gamma$-band levels
are known only up to $J^\pi=9^+_\gamma$ in $^{126}$Xe
and up to $J^\pi=7^+_\gamma$ in $^{128}$Xe.
It would be of interest to confirm experimentally
the predicted signature splitting for the higher-spin states.

The $\gamma$ bands in the Ba isotopes are peculiar
in the sense that their observed even--odd staggering
is {\em larger} than the one calculated without cubic interaction.
Again, this deviation increases with angular momentum $J$
and can be corrected with the $\hat B_3^\dag\cdot\tilde B_3$ interaction
but this time with a {\em positive} coefficient $v_3$
(see table~\ref{t_parxeba}).
In fact, for $J\geq8$ the signature splitting $|S(J)|$
is larger than 1 in $^{128}$Ba and $^{132}$Ba
which corresponds to a $7^+_\gamma$ above $8^+_\gamma$
and a $9^+_\gamma$ above $10^+_\gamma$.
This peculiar behaviour can be modelled to some extent
with a cubic interaction.

\begin{figure}
\centering
\includegraphics[width=6.5cm]{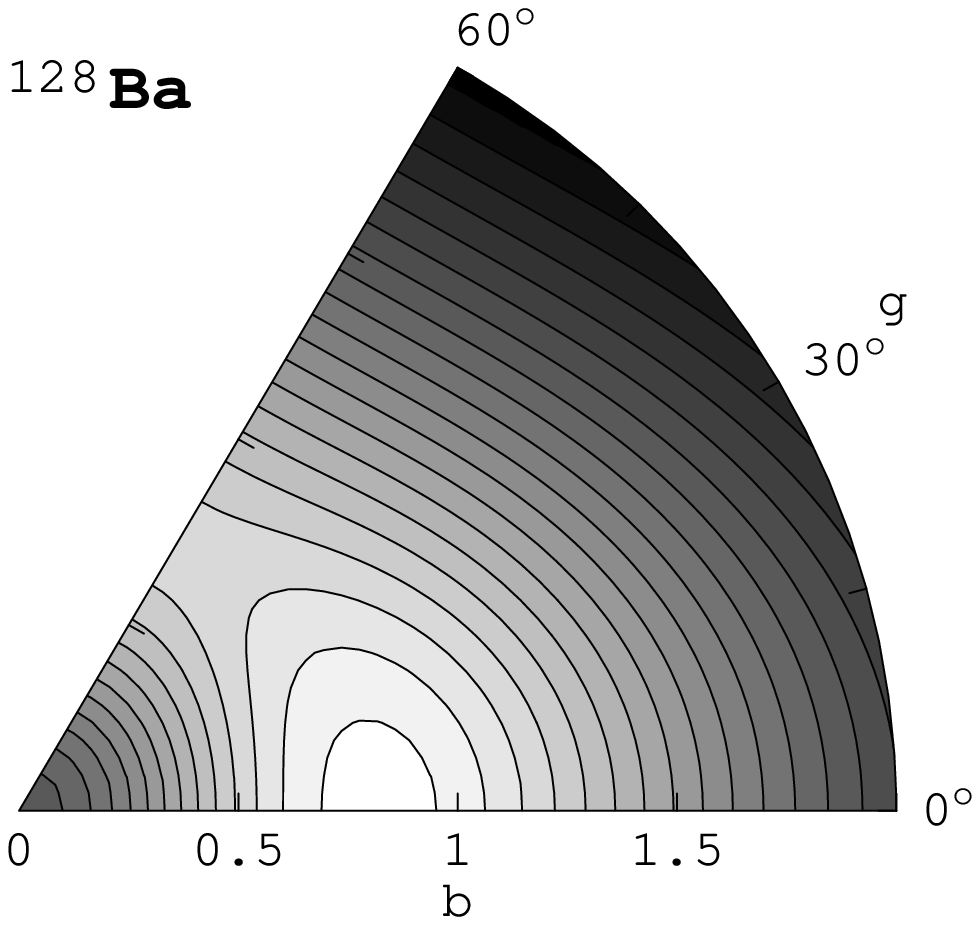}
\includegraphics[width=6.5cm]{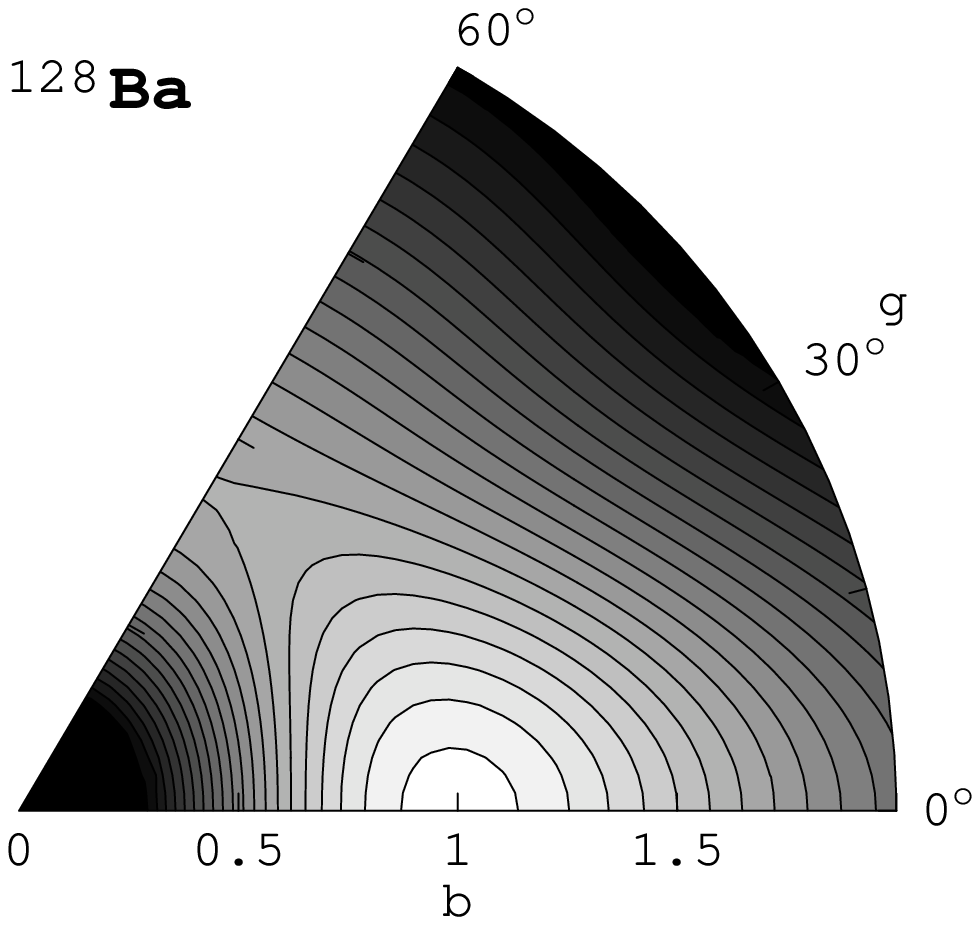}
\caption{
Same caption as fig.~\ref{f_rup} for the nucleus $^{128}$Ba.}
\label{f_ba128p}
\end{figure}

The positive value of the fitted coefficient $v_3$
has the expected effect on the potential $V(\beta,\gamma)$
derived in the classical limit,
as is illustrated in fig.~\ref{f_ba128p} for the nucleus $^{128}$Ba.
Whereas a negative $v_3$ leads to a softer potential
in the $\gamma$ direction (as in $^{112}$Ru),
a positive value for this interaction yields a more rigid, axially symmetric shape
with a slightly larger deformation $\beta$.

In the initial two-body hamiltonian
the E2 transition rates depend strongly on the value of $\chi$
in the quadrupole operator.
It is expected that this is still the case
when cubic terms are added to the hamiltonian
as long as these do not substantially alter its eigenstates.
In several of the Xe and Ba isotopes
many $B$(E2) values between the low-lying states are known
and these allow a test of the wave function in the cubic calculation.
The results are shown in tables~\ref{t_e2xe} and~\ref{t_e2ba}.
\begin{table}
\caption{
Experimental and calculated $B$(E2) values
in units of $10^2$ e$^2$fm$^4$ in $^{124-128}$Xe.}
\label{t_e2xe}
\vspace{3mm}
\begin{tabular}{cccccccccc}
\hline
&~~~&\multicolumn{2}{c}{$^{124}$Xe}
&~~~&\multicolumn{2}{c}{$^{126}$Xe}
&~~~&\multicolumn{2}{c}{$^{128}$Xe}\\
\cline{3-4}\cline{6-7}\cline{9-10}
$J^\pi_{\rm i}\rightarrow J^\pi_{\rm f}$
&&Expt&IBM-1c&&Expt&IBM-1c&&Expt&IBM-1c\\
\hline
$2_1^+\rightarrow 0_1^+$&&30(2)&30&&15(5)&15&&15(1)&18\\
$4_1^+\rightarrow 2_1^+$&&40(5)&42&&---&24&&23(2)&24\\
$6_1^+\rightarrow 4_1^+$&&48(18)&45&&---&29&&30(3)&24\\
$8_1^+\rightarrow 6_1^+$&&19(8)&43&&---&30&&37(4)&20\\
$2_2^+\rightarrow 0_1^+$&&0.48(11)&0.65&&---&0.00&&0.25(2)&0.20\\
$2_2^+\rightarrow 2_1^+$&&21(4)&33&&---&24&&18(2)&21\\
$3_1^+\rightarrow 2_1^+$&&0.32(12)&0.80&&---&0.00&&---&0.21\\
$3_1^+\rightarrow 2_2^+$&&2.9(11)&34&&---&22&&---&18\\
$4_2^+\rightarrow 2_2^+$&&26(9)&22&&---&15&&---&12\\
\hline
\end{tabular}
\end{table}
\begin{table}
\caption{
Experimental and calculated $B$(E2) values
in units of $10^2$ e$^2$fm$^4$ in $^{128-132}$Ba.}
\label{t_e2ba}
\vspace{3mm}
\begin{tabular}{cccccccccc}
\hline
&~~~&\multicolumn{2}{c}{$^{128}$Ba}
&~~~&\multicolumn{2}{c}{$^{130}$Ba}
&~~~&\multicolumn{2}{c}{$^{132}$Ba}\\
\cline{3-4}\cline{6-7}\cline{9-10}
$J^\pi_{\rm i}\rightarrow J^\pi_{\rm f}$
&&Expt&IBM-1c&&Expt&IBM-1c&&Expt&IBM-1c\\
\hline
$2_1^+\rightarrow 0_1^+$&&28(3)&29&&22(1)&22&&17(2)&26\\
$4_1^+\rightarrow 2_1^+$&&41(2)&39&&31(1)&31&&---&40\\
$6_1^+\rightarrow 4_1^+$&&39(3)&40&&37(2)&31&&---&44\\
$8_1^+\rightarrow 6_1^+$&&37(5)&37&&35(1)&28&&---&39\\
$2_2^+\rightarrow 0_1^+$&&1.3(2)&1.2&&---&1.1&&1.5(2)&0.17\\
$2_2^+\rightarrow 2_1^+$&&---&24&&---&17&&57(6)&37\\
$4_2^+\rightarrow 2_1^+$&&0.34(2)&0.25&&---&0.47&&---&0.27\\
$4_2^+\rightarrow 2_2^+$&&24(2)&24&&---&17&&---&23\\
$6_2^+\rightarrow 4_1^+$&&0.30(4)&0.40&&---&0.23&&---&0.39\\
$6_2^+\rightarrow 4_2^+$&&38(5)&28&&---&21&&---&27\\
\hline
\end{tabular}
\end{table}
Generally a good agreement
between experimental and calculated $B$(E2) values is obtained.
One notable discrepancy
is the $2_2^+\rightarrow 0_1^+$ transition in $^{132}$Ba
with a calculated $B$(E2) value which is an order of magnitude too small.
This value is equally small in the \mbox{IBM-1} calculation without cubic interaction
and is due to an accidental cancellation of terms
with the hamiltonian~(\ref{e_ham12}).

\subsection{Osmium and platinum isotopes}
\label{ss_ospt}
Cizewski {\it et al.}~\cite{Cizewski78} proposed $^{196}$Pt
as a first example of the SO(6) dynamical symmetry of the \mbox{IBM-1}.
Subsequently, it became clear that the entire region of Pt and Os isotopes
can be described as transitional between the SO(6) and SU(3) limits~\cite{Casten78}.
In this subsection we present a detailed analysis of Os and Pt isotopes
of which enough $\gamma$-band levels are firmly established
for the staggering analysis to be meaningful.
In both $^{186}$Os and $^{186}$Pt the $\gamma$-band levels
are known up $J^\pi=10^+$ such that its staggering properties
can be fitted reliably in these nuclei.
Less information is available in $^{180}$Os and $^{182}$Os (up to $7^+_\gamma$),
and even less in $^{184}$Os (up to $6^+_\gamma$).
The latter nucleus, although relatively poorly known, is included in the analysis
since it is intermediate between two better known isotopes.

\begin{figure}
\centering
\includegraphics[width=9cm]{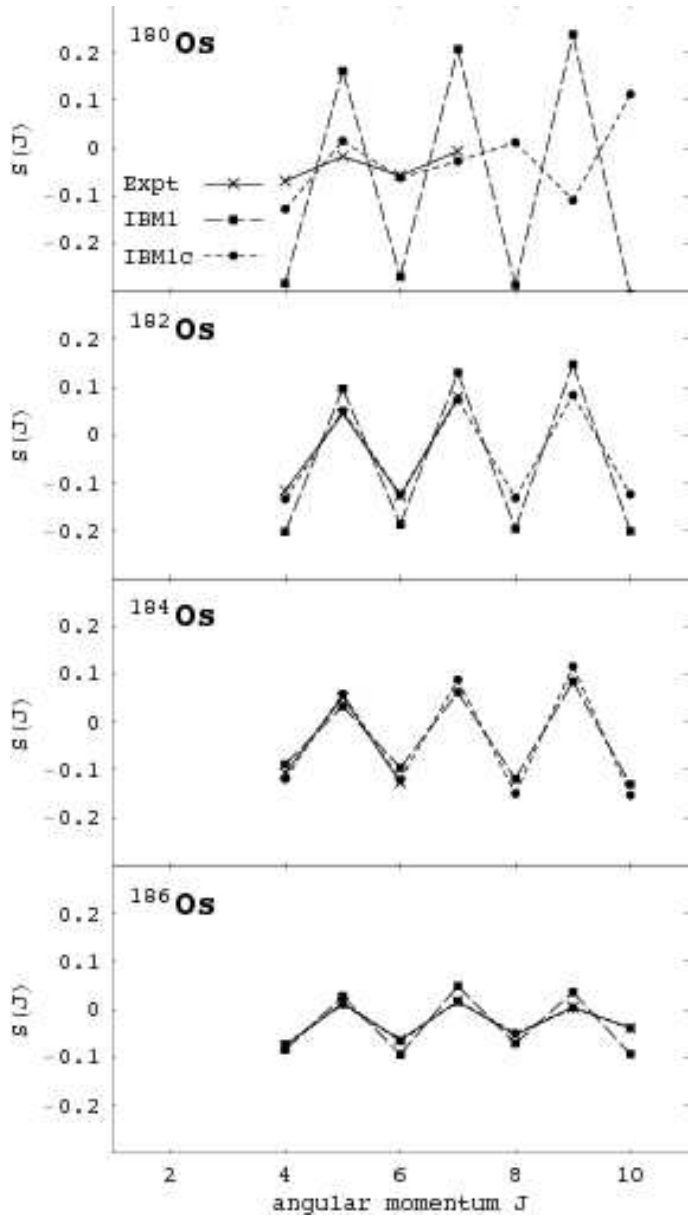}
\caption{
Same caption as fig.~\ref{f_rus} for the nuclei $^{180-186}$Os.}
\label{f_oss}
\end{figure}

\begin{figure}
\centering
\includegraphics[width=9cm]{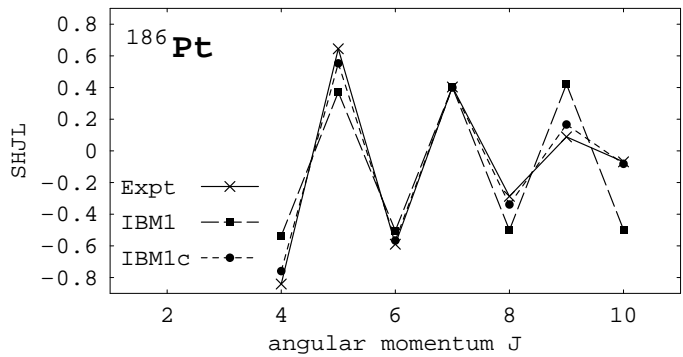}
\caption{
Same caption as fig.~\ref{f_rus} for the nucleus $^{186}$Pt.}
\label{f_pt186s}
\end{figure}

The parameters resulting from the fits are shown in table~\ref{t_parospt}.
\begin{table}
\caption{
Parameters and rms deviation for Os and Pt isotopes in units of keV.}
\label{t_parospt}
\vspace{3mm}
\begin{tabular}{ccccccccccccccc}
\hline
Nucleus&~&
$\epsilon_d$&~&$\kappa$&~&$\kappa'$&~&$\lambda_d$&~&$v_3$&~&$\chi^*$
&~&$\sigma$\\
\hline
$^{180}$Os
&&$~946$&&$-29.1$&&$~~4.1$&&$-62.5$&&---&&$-0.30$&&$19$\\
&&$~634$&&$-41.2$&&$-5.1$&&$-12.0$&&$-58.7$&&$-0.30$&&$10$\\
$^{182}$Os
&&$1013$&&$-32.7$&&$~~5.7$&&$-81.3$&&---&&$-0.30$&&$~8$\\
&&$~813$&&$-37.8$&&$~~2.8$&&$-62.3$&&$-14.9$&&$-0.30$&&$~9$\\
$^{184}$Os
&&$~844$&&$-40.6$&&$~~6.8$&&$-89.7$&&---&&$-0.30$&&$~2$\\
&&$~973$&&$-35.6$&&$~~9.3$&&$-99.2$&&$~~10.0$&&$-0.30$&&$~1$\\
$^{186}$Os
&&$~871$&&$-34.8$&&$~10.9$&&$-95.7$&&---&&$-0.30$&&$~5$\\
&&$~752$&&$-40.4$&&$~~7.8$&&$-84.5$&&$-16.1$&&$-0.30$&&$~1$\\
\hline
$^{186}$Pt
&&$~747$&&$-33.7$&&$~~3.7$&&$-65.9$&&---&&$-0.20$&&$36$\\
&&$~449$&&$-15.2$&&$-3.9$&&~~$15.5$&&$-60.5$&&$-0.20$&&$24$\\
\hline
\multicolumn{15}{l}{$^*$Dimensionless.}
\end{tabular}
\end{table}
Since the Os isotopes are further removed from SO(6) and closer to SU(3),
the value of $|\chi|$ is larger than in the other nuclei considered so far.
The more deformed character of the Os isotopes
also leads to a parameter systematics
which is smoother than in previous examples.
It should be noted, however, that in $^{184}$Os
the sign of $v_3$ comes out positive
while in the other isotopes it is negative.
This conceivably might be due to the fact that not enough levels are known
in the $\gamma$ band of this nucleus.
The main outcome of the fits to $^{184}$Os and $^{186}$Os
is that the $\hat B_3^\dag\cdot\tilde B_3$ interaction is small.
The results for the $\gamma$-band staggering
in $^{180-186}$Os and in $^{186}$Pt
are shown in figs.~\ref{f_oss} and~\ref{f_pt186s}, respectively.
Given the low rms deviations obtained in the fit,
it comes as no surprise that in all nuclei 
the observed signature splitting of the $\gamma$ band
is accurately reproduced.
It is also seen that the inclusion of the $\hat B_3^\dag\cdot\tilde B_3$ interaction
leads to a systematic improvement in the description of the $\gamma$ band.
A noteworthy example is $^{186}$Os.
The standard two-body \mbox{IBM-1} hamiltonian
already gives an excellent description with an rms deviation of only $\sigma=5$~keV.
Nevertheless, there is a small but consistent deviation
in the staggering pattern of the $\gamma$ band
which is remedied with the cubic term.
The nucleus $^{186}$Os is also well studied as regards its E2 decay properties
and, as can be seen from table~\ref{t_e2os},
all its known $B$(E2) values are rather well reproduced by the calculation.
\begin{table}
\caption{
Experimental and calculated $B$(E2) values
in units of $10^2$ e$^2$fm$^4$ in $^{180-186}$Os.}
\label{t_e2os}
\vspace{3mm}
\begin{tabular}{ccccccccccc}
\hline
&~&\multicolumn{2}{c}{$^{180}$Os}
&~&\multicolumn{1}{c}{$^{182}$Os}
&~&\multicolumn{1}{c}{$^{184}$Os}
&~&\multicolumn{2}{c}{$^{186}$Os}\\
\cline{3-4}\cline{6-6}\cline{8-8}\cline{10-11}
$J^\pi_{\rm i}\rightarrow J^\pi_{\rm f}$
&&Expt&IBM-1c&&IBM-1c&&IBM-1c&&Expt&IBM-1c\\
\hline
$2_1^+\rightarrow 0_1^+$&&76(19)&74&&77&&61&&58(2)&69\\
$4_1^+\rightarrow 2_1^+$&&122(16)&107&&110&&86&&85(4)&97\\
$6_1^+\rightarrow 4_1^+$&&101(25)&119&&120&&90&&116(3)&104\\
$8_1^+\rightarrow 6_1^+$&&40(8)&125&&124&&88&&110(6)&103\\
$2_2^+\rightarrow 0_1^+$&&---&4.3&&3.9&&2.4&&6.4(3)&3.1\\
$2_2^+\rightarrow 2_1^+$&&---&17&&11&&4.6&&14.8(5)&8.6\\
$2_2^+\rightarrow 4_1^+$&&---&1.3&&0.72&&0.32&&0.8(3)&0.60\\
$4_2^+\rightarrow 2_1^+$&&---&0.37&&0.80&&1.0&&2.0(2)&0.70\\
$4_2^+\rightarrow 4_1^+$&&---&17&&11&&4.2&&15.6(13)&8.3\\
$4_2^+\rightarrow 2_2^+$&&---&41&&42&&33.0&&45(4)&34\\
$6_2^+\rightarrow 4_1^+$&&---&0.00&&0.26&&0.79&&0.80(7)&0.24\\
$6_2^+\rightarrow 4_2^+$&&---&77&&79&&60&&75(7)&62\\
\hline
\end{tabular}
\end{table}
In two of the four isotopes, namely $^{182}$Os and $^{184}$Os,
only one or two $B$(E2) values are known experimentally,
and for these nuclei only theoretical predictions are quoted in the table.

\begin{figure}
\centering
\includegraphics[width=6.5cm]{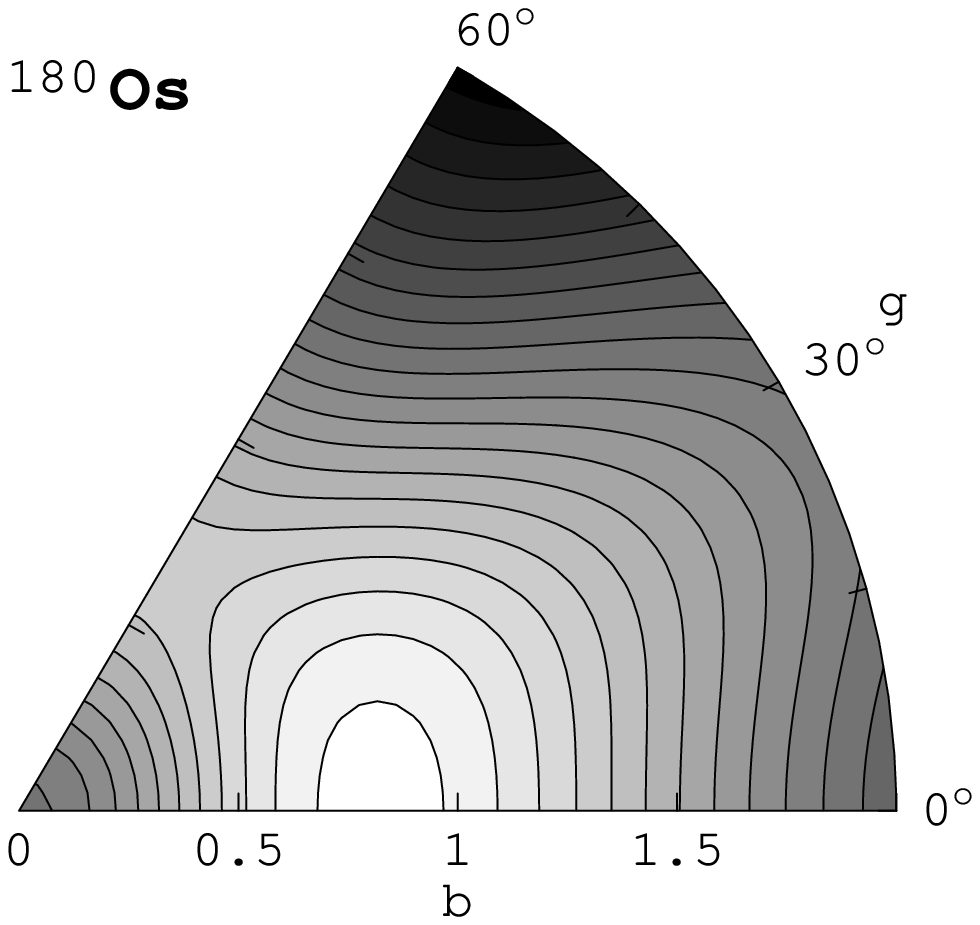}
\includegraphics[width=6.5cm]{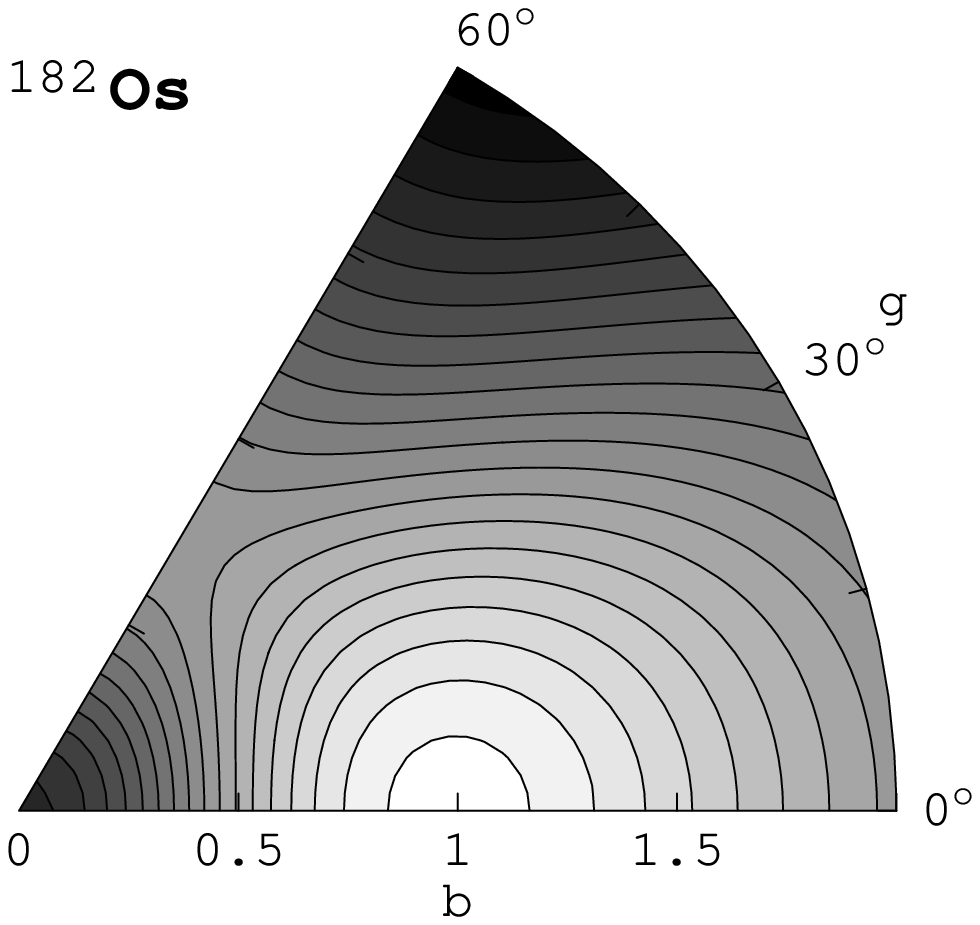}
\includegraphics[width=6.5cm]{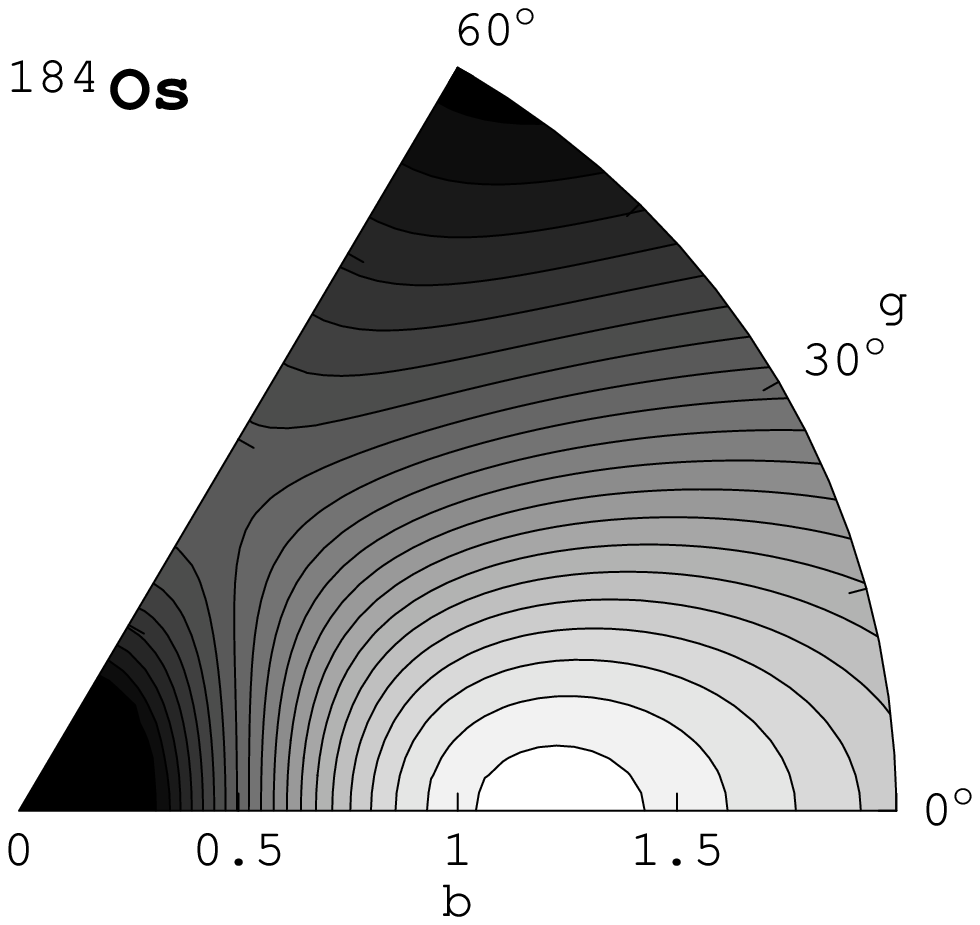}
\includegraphics[width=6.5cm]{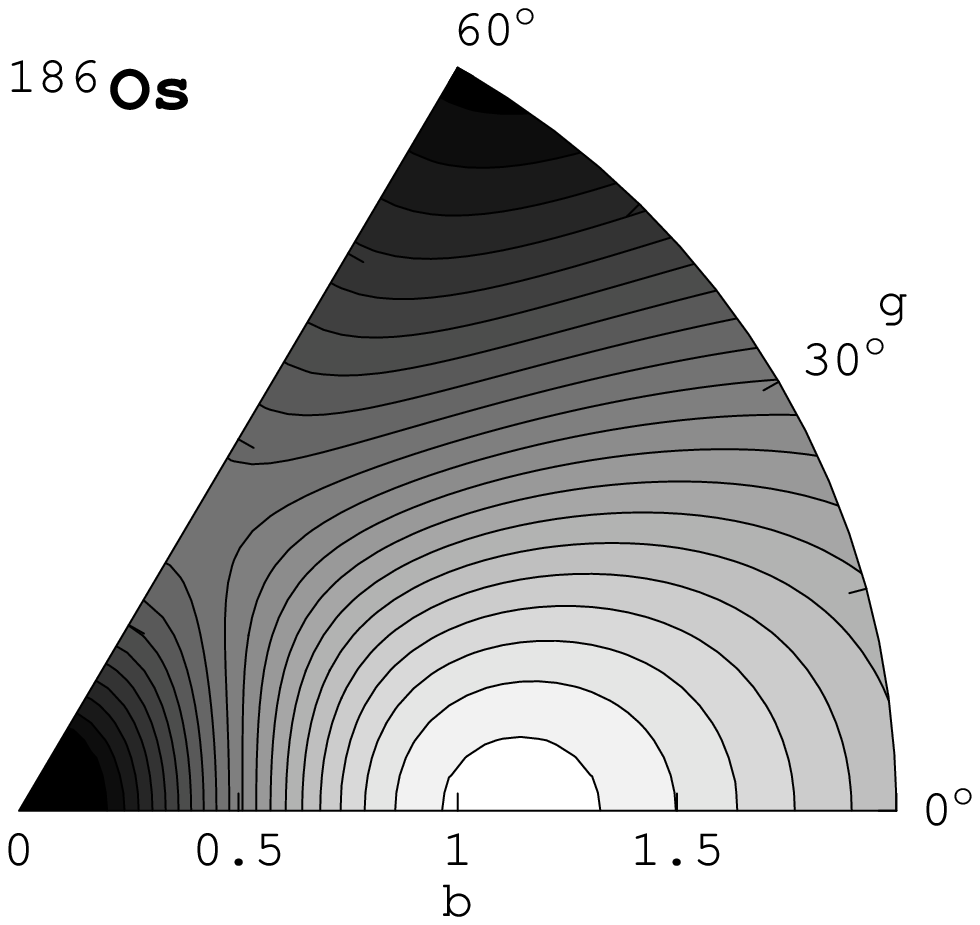}
\caption{
Potential energy surfaces $V(\beta,\gamma)$ for the nuclei $^{180-186}$Os.
All plots show the classical limit of the \mbox{IBM-1} hamiltonian
with the cubic interaction included.}
\label{f_osptp}
\end{figure}

As a final outcome of this study we show in fig.~\ref{f_osptp}
the potential energy surfaces for the different Os isotopes
which are obtained in the classical limit of the \mbox{IBM-1} hamiltonian
with cubic interactions.
It shows the evolution with increasing mass number $A$
towards a potential with a more deformed minimum
and which is more rigid in $\gamma$.

\section{Conclusions}
\label{s_con}
Two main conclusions can be drawn from this work.
First, very accurate nucleus-by-nucleus fits
can be achieved with the simplified \mbox{IBM-1} hamiltonian~(\ref{e_ham12})
to which a single three-body interaction of the type~(\ref{e_ham3}) is added.
In almost all $\gamma$-soft nuclei studied
the inclusion of the $\hat B_3^\dag\cdot\tilde B_3$ interaction
yields a consistently better description
of the signature splitting of the $\gamma$ band.
This improved description of the $\gamma$-band energies is obtained
while maintaining good agreement for the E2-decay properties.
Nevertheless, no systematics of three-body parameters could be established.
While we have currently a good working hamiltonian
which includes up to two-body interactions
and which describes nuclei throughout the nuclear chart,
little is known of the overall trends for three-body interactions.

A second conclusion concerns the geometry
underlying the algebraic hamiltonians,
as was illustrated with several examples.
One surprising outcome of our approach
is that an unbiased fit of energy levels in some nuclei
($^{114,116}$Pd, $^{126}$Xe, $^{132}$Ba and $^{186}$Pt)
leads to an \mbox{IBM-1} hamiltonian
of which the classical limit yields a potential with a spherical minimum
whereas these nuclei usually are considered $\gamma$ soft and weakly deformed.
The additional three-body interaction $v_3\hat B_3^\dag\cdot\tilde B_3$,
introduced to improve the description of the $\gamma$ band,
makes the potential energy surface $V(\beta,\gamma)$ softer in $\gamma$ for $v_3<0$
and more rigid for $v_3>0$.
The sign of this coefficient follows from the staggering pattern of the $\gamma$ band.
Finally, in none of the nuclei studied we found evidence for a triaxial minimum.

\section*{Acknowledgement}
We wish to thank Adrian Gelberg for his valuable comments on this work.

\end{document}